\documentclass[twocolumn]{aastex62}
\usepackage{natbib}
\usepackage{threeparttable}
\usepackage{amssymb}
\usepackage{amsmath}
\usepackage{graphicx}
\usepackage{CJKutf8}
\usepackage{textcomp}

\shorttitle{Fast and slow paths to quiescence}
\shortauthors{Wu et al.}

\begin{document}
\title{Fast and slow paths to quiescence: ages and sizes of 400 quiescent galaxies from the LEGA-C survey}

\author{Po-Feng Wu \begin{CJK*}{UTF8}{bkai}(吳柏鋒)\end{CJK*}}
\affiliation{Max-Planck-Institut f\"{u}r Astronomie, K\"{o}nigstuhl 17, D-69117, Heidelberg, Germany}

\author{Arjen van der Wel}
\affiliation{Sterrenkundig Observatorium, Universiteit Gent, Krijgslaan 281 S9, B-9000 Gent, Belgium}
\affiliation{Max-Planck-Institut f\"{u}r Astronomie, K\"{o}nigstuhl 17, D-69117, Heidelberg, Germany}

\author{Rachel Bezanson}
\affiliation{University of Pittsburgh, Department of Physics and Astronomy, 100 Allen Hall, 3941 O'Hara St, Pittsburgh PA 15260, USA}

\author{Anna Gallazzi}
\affiliation{INAF-Osservatorio Astrofisico di Arcetri, Largo Enrico, Fermi 5, I-50125 Firenze, Italy}

\author{Camilla Pacifici}
\affiliation{Space Telescope Science Institute, 3700 San Martin Drive, Baltimore, MD 21218, USA}

\author{Caroline M. S. Straatman}
\affiliation{Sterrenkundig Observatorium, Universiteit Gent, Krijgslaan 281 S9, B-9000 Gent, Belgium}

\author{Ivana Bari\v{s}i\'{c}}
\affiliation{Max-Planck-Institut f\"{u}r Astronomie, K\"{o}nigstuhl 17, D-69117, Heidelberg, Germany}

\author{Eric F. Bell}
\affiliation{Department of Astronomy, University of Michigan, 1085 South University Avenue, Ann Arbor, MI 48109-1107, USA}

\author{Priscilla Chauke}
\affiliation{Max-Planck-Institut f\"{u}r Astronomie, K\"{o}nigstuhl 17, D-69117, Heidelberg, Germany}

\author{Josha van Houdt}
\affiliation{Max-Planck-Institut f\"{u}r Astronomie, K\"{o}nigstuhl 17, D-69117, Heidelberg, Germany}

\author{Marijn Franx}
\affiliation{Leiden Observatory, Leiden University, PO Box 9513, 2300 RA Leiden, The Netherlands}

\author{Adam Muzzin}
\affiliation{Department of Physics and Astronomy, York University, 4700 Keele St., Toronto, Ontario, M3J 1P3, Canada}

\author{David Sobral}
\affiliation{Physics Department, Lancaster University, Lancaster LA1 4YB, UK}
\affiliation{Leiden Observatory, Leiden University, PO Box 9513, 2300 RA Leiden, The Netherlands}

\author{Vivienne Wild}
\affiliation{School of Physics and Astronomy, University of St Andrews, North Haugh, St Andrews, KY16 9SS, U.K.}

\correspondingauthor{Po-Feng Wu}
\email{pofeng@mpia.de}

\begin{abstract}	
We analyze stellar age indicators (D$_n$4000 and EW(H$\delta$)) and sizes of 467 quiescent galaxies with $M_\ast \geq 10^{10} M_\odot$ at $z\sim0.7$ drawn from DR2 of the LEGA-C survey. Interpreting index variations in terms of equivalent single stellar population age, we find that the median stellar population is younger for larger galaxies at fixed stellar mass. The effect is significant, yet small; the ages of the larger and the smaller subsets differ by only $<500$~Myr, much less than the age variation among individual galaxies ($\sim1.5$~Gyr). At the same time, post-starburst galaxies --- those experienced recent and rapid quenching events --- are much smaller than expected based on the global correlation between age and size of normal quiescent galaxies. These co-existing trends unify seemingly contradictory results in the literature; the complex correlations between size and age indicators revealed by our large sample of galaxies with high-quality spectra suggest that there are multiple evolutionary pathways to quiescence. Regardless of the specific physical mechanisms responsible for the cessation of star formation in massive galaxies, the large scatter in D$_n$4000 and EW(H$\delta$) immediately implies that galaxies follow a large variety in evolutionary pathways.
On the one hand, we see evidence for a process that slowly shuts off star-formation and transforms star-forming galaxies to quiescent galaxies without necessarily changing their structures. On the other hand, there is likely a mechanism that rapidly quenches galaxies, an event that coincides with dramatic structural changes, producing post-starburst galaxies that can be smaller than their progenitors. 

\end{abstract}


\section{Introduction}

Large extragalactic surveys have revealed that there are two distinct populations of galaxies: the so-called red sequence, dominated by galaxies with quiescent star formation and old stellar populations, and the blue cloud, containing mainly star-forming galaxies \citep{str01,bal04,bel04,wil06,fra07}. The structure of stellar components of these two categories of galaxies shows clear differences. Quiescent galaxies have on average smaller sizes and more concentrated light profiles than star-forming galaxies of the same stellar masses. The average sizes of each population increase by several times from $z\sim2$ to $z\sim0$ and the difference in size is in place at all epochs observed \citep{she03,tru07,wil10,vdw14,pau17}.
Despite the empirical correlation between galaxy sizes and star-formation rates (SFRs) being well established, the physical causality \citep[or lack of;][]{lil16} remains contentious.

Several mechanisms have been proposed to explain the cessation of star-formation in massive galaxies. For field galaxies, 
one type of mechanisms is cutting off the cold gas supply, leading to galaxies naturally running out of fuel for forming new stars. In massive dark matter halos, infalling gas would be shock heated to high temperature \citep{dek06}. Additionally, active galactic nuclei (AGN) or supernovae could inject energy and heat up the gas, further slowing down the cooling process and star-formation \citep{bes05,cro06,fab12,bar17,ter17}. The star-formation may continue for a few Gyrs until the cold gas reservoir is used up or heated up. 

Another type involves more violent events. Gas-rich galaxy mergers, interaction, and disk instability can efficiently funnel gas into the centers of galaxies, induce intensive star formation and exhaust the available gas in a short period of time \citep{bar91,bar96,sny11}. Gas can be also removed from the galaxies due to strong outflows induced by starbursts and AGNs \citep{spr05,kav07}. Furthermore, the turbulence induced by the violent processes prevents the remaining gas from collapsing and forming stars \citep{ell18,sme18}.

These two types of processes predict different structural change during the transition period. The first type of mechanisms does not directly involve a change of galaxy structure; therefore, new quiescent galaxies will have sizes similar to their star-forming progenitors. The second category involves further star-formation activities in the galaxy centers. Dense cores build up during the process and the structures change. 

The expected sizes of newly-formed quiescent galaxies are thus different in these two scenarios. If the sizes do not change much during the transition from star-forming to quiescent, newly-formed quiescent galaxies are expected to be larger than the existing quiescent population. At any epoch, there should be a general correlation between the stellar ages and the sizes of quiescent galaxies such that at fixed stellar mass, larger galaxies are on average younger. This formation process also provides, at least partially, a natural explanation to the smaller average sizes of quiescent galaxies in the early universe. The observed size evolution is a `progenitor bias' in such that large quiescent galaxies at present-day were still forming stars at higher redshifts, thus, not classified as quiescent galaxies. The population of quiescent galaxies at high-redshifts is a smaller biased subset of present-day quiescent galaxies \citep{vd01,vdw09,car13,pog13}. 
On the other hand, if the quenching process is accompanied by the growth of a dense galaxy core, the effective radii of newly-formed quiescent galaxies become smaller than their progenitors and can be even smaller than existing quiescent galaxies. 

The picture is complicated by subsequent merger events. Dry mergers add stars of any ages to the outskirts of quiescent galaxies, therefore, the sizes of quiescent galaxies increase and the average age of the stellar population is altered. The age-size correlation, if existing at the time when quiescent galaxies form, will be gradually washed out by the merger events \citep{sha10a}. This issue can be mitigated by looking at the high-redshift universe, where galaxies are on average much younger, such that fewer merger events had happened by the time that galaxies are observed. 

Some recent works have attempted to address the age-size correlation up to $z\sim2$ but the results are so far inconclusive. Studies splitting the population by sizes generally report either weak or no significant correlation that larger galaxies are younger \citep{vdw09,tru11,fag16,zan16,wil17}. On the contrary, multiple works find that post-starburst (PSB) galaxies, those young quiescent galaxies whose star-formation rates dropped rapidly in the recent past \citep{dre83,bal99,dre99}, are smaller than the average quiescent galaxies \citep{whi12,yan16,alm17}. 
These two statements appear to be supporting different scenarios of the formation and the size evolution of quiescent galaxies. 

At the moment, it is not straight-forward to combine all these results and form a coherent picture. Each work studies different galaxy samples in various stellar mass and redshift ranges, measures the galaxy sizes and ages from imaging and spectroscopic data of different qualities with various methods. 
In this paper, we present an overview of the sizes and age-sensitive absorption features (D$_n$4000 and EW(H$\delta$)) of over 400 quiescent galaxies. The sample is drawn from the Large Early Galaxy Astrophysics Census (LEGA-C) survey \citep[][Straatmann et al. submitted]{vdw16}. All galaxies have ultra-deep optical spectra that provide precise measurements to infer the ages and recent star-formation histories \citep{wu18b,cha18} as well as images from the \textit{Hubble Space Telescope} (HST) for accurate size measurements. 

We will show that the correlations between the size and age indicators is complex and can only be revealed by a large sample of galaxies with high-quality spectra. Our results suggest that galaxies join the red sequence in different ways.

We describe the galaxy sample and the measurement of basic galaxy properties in Section~\ref{sec:data}. 
In Section~\ref{sec:res}, we investigate the correlation between ages and sizes on an individual-galaxy basis. In Section~\ref{sec:dis} we briefly summarize the literature and discuss the implications for the formation of quiescent galaxies. We present our conclusions in Section~\ref{sec:con}.

\section{Data and Analysis}
\label{sec:data}
This study is based on the first two years of data of the LEGA-C survey. The LEGA-C survey is a 4-year survey using the Visible Multi-Object Spectrograph \citep[VIMOS;][]{lef03} mounted on the 8 m Very Large Telescope to obtain rest-frame optical spectra of $\sim$3000 $K_s$-band selected galaxies mainly at $0.6 \leq z \leq 1.0$. Each galaxy receives $\sim20$ hrs of integration at a spectral resolution of $R\sim3500$. The typical continuum signal-to-noise ratio (S/N) is 20\AA$^{-1}$. We refer to \citet{vdw16} for the design of the survey and Straatman et al. (2018) for details of observation and data reduction.

\subsection{Stellar Masses, star-formation rates, and sizes of galaxies}
\label{sec:mr}

We derive galaxy stellar masses by fitting the observed multi-wavelength spectral energy distributions (SEDs) from the UltraVISTA catalog \citep{muz13} using the FAST code \citep{kri09}. The SED templates are from the \citep{bc03} stellar population synthesis models with exponentially declining star-formation rates. We adopt a \citet{cha03} initial mass function (IMF) and the \citet{cal00} dust extinction law. The SFRs are estimated from the UV and IR luminosities, following the prescription of \citet{whi12}.

We derive the sizes of galaxies using the \textit{HST} ACS F814W images from the COSMOS program \citep[GO-9822, GO-10092, PI: N. Scoville, ][]{sco07}, following the procedure of \citet{vdw12}. For each galaxy, we first create a 10\arcsec\ cutout and then use \texttt{galfit} \citep{pen10} to fit a single S\'{e}rsic profiles to the 2-D light profile. The effective radius, total magnitude, S\'{e}rsic index, axis ratio, position angle, and background are set as free parameters. Neighboring sources are fit simultaneously. The Point Spread Function (PSF) is measured from stars in the proximity in terms of the detector position. The size of galaxies, $R_e$, is defined as the semi-major axis of the ellipse that contains half of the total flux of the best-fit S\'{e}rsic model.

\subsection{Measuring spectral indices}

In this paper, we examine two age-sensitive spectral features, the 4000\AA\ break, D$_n$4000 and the equivalent width of H$\delta$ absorption, EW(H$\delta$). We use the equivalent width of H$\beta$ emission to distinguish quiescent from star-forming galaxies. 

To separate the ionized gas emission from the stellar continuum, we model the observed spectrum using the Penalized Pixel-Fitting (pPXF) method \citep{cap04,cap17}. Each galaxy spectrum is fit by a combination of two templates representing the stellar continuum and the gas emission. We model the continuum with high resolution ($R=10,000$) theoretical single stellar population (SSP) templates (Conroy et al., in prep.). All emission lines are modeled as a single kinematic component: all with the same velocity and velocity dispersion. The strength of each line is a free parameter. We refer to \citet{bez18} for the detailed fitting process. 

We adopt the definition of the D$_n$4000 in \citet{bal99} and the H$\delta_a$ index in \citet{wor97}. Both indices are measured from the emission-line-subtracted spectra. The equivalent width of H$\beta$ (EW(H$\beta$)), is the ratio between the best-fit model emission line flux and the local continuum level, which is defined as the average continuum flux density between 4760\AA\ and 4960\AA\ in rest-frame. The typical uncertainties are $\sim0.025$, $\sim0.40$\AA, and $\sim0.35$\AA\ for D$_n$4000, EW(H$\delta$), and EW(H$\beta$), respectively.

\subsection{Sample Selection}

\begin{figure*}
	\centering
	\includegraphics[width=0.9\textwidth]{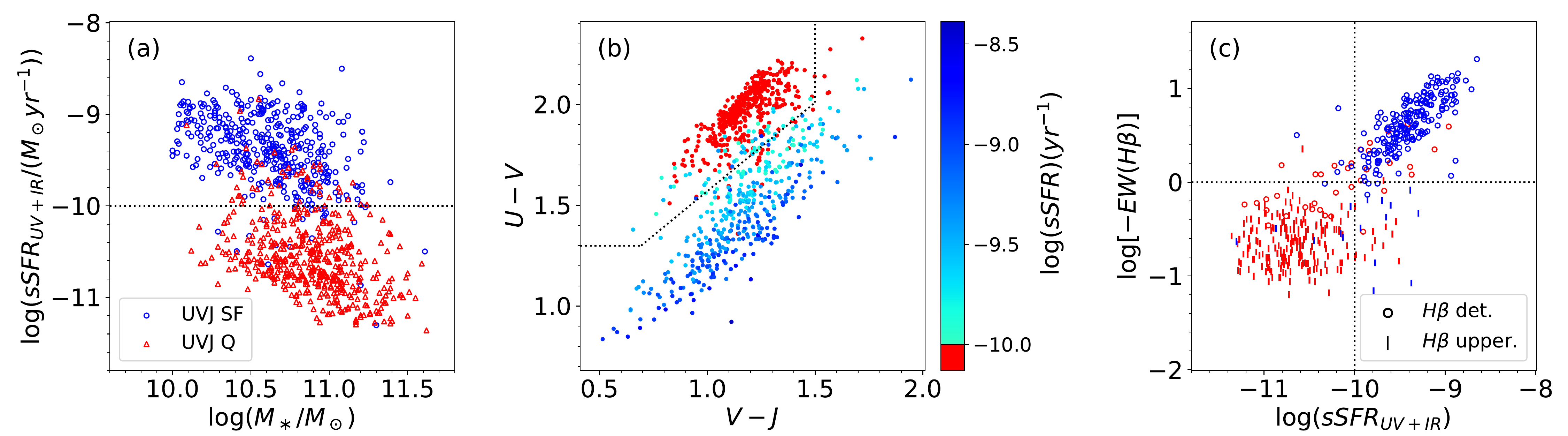}
	\caption{Comparison of 3 different methods separating star-forming and quiescent galaxies: the equivalent width of H$\beta$ emission line, sSFR derived from UV and IR fluxes, and the UVJ two-color scheme. (a) The sSFR as a function of stellar masses. We adopt a fiducial sSFR limit of $10^{-10} M_\odot\ \mbox{yr}^{-1}$ separating star-forming and quiescent galaxies (the dotted horizontal line). Blue circles and red triangles are star-forming and quiescent galaxies based on the UVJ two-color scheme. (b) The $U-V$ v.s. $V-J$ two-color diagram. Galaxies are color-coded according to the sSFR. The dotted line labels the demarcation separating star-forming from quiescent galaxies of \citet{muz13}. Galaxies at the top-left corner are classified as quiescent galaxies based on the two-color scheme. (c) EW(H$\beta$) as a function of sSFR. The circles are galaxies whose H$\beta$ emission is detected. The short vertical lines are $2\sigma$ upper-limits of non-detections. The vertical and horizontal dotted lines are the sSFR and EW(H$\beta$) demarcations we adopted in the paper. Blue and red data points are star-forming and quiescent galaxies based on the UVJ two-color scheme. Panel (c) shows that the 3 different methods, in general, agree with each other well.}
	\label{fig:sfq}
\end{figure*}

The first two years of LEGA-C primary sample consists of 1550 galaxies brighter than $K_s = 20.7 - 7.5 \times \log[(1+z)/1.8]$ and with redshifts $0.6 \leq z \leq 1.0$ \citep{vdw16}. We exclude spectra (1) with fundamental flaws, (2) that we cannot measure redshifts from, and (3) whose pPXF fit is flawed. The remaining 1462 spectra are considered to be useful for scientific purposes and have a flag $f_{use}=1$ in the LEGA-C DR2 catalog (Straatmann et al. 2018, submitted). 

To measure the EW(H$\delta$), we first select 1112 spectra that cover the wavelength range of EW(H$\delta$) and with median $S/N \geq 10$ per spectral pixel between rest-frame wavelength 4000\AA\ and 4300\AA. The uncertainties on EW(H$\delta$) of spectra with $S/N<10$ are in general $>1$\AA, too large to be a robust age indicator and would contaminate the selection of PSB galaxies. 
We then visually inspect \textit{HST} and the \texttt{galfit} best-fit model images, exclude 159 galaxies with catastrophic failures in the fitting or highly disturbed morphologies that cannot be well modeled by a single S\'{e}rsic profile. From the remaining 953 galaxies, we further exclude 61 galaxies whose spectra or broadband photometries are contaminated by neighboring objects. In total, we have 892 galaxies with good measurements of spectral indices, sizes, and stellar masses. 

Several methods can be used to select quiescent populations, resulting in different levels of contamination and incompleteness \citep{mor13}. In this paper, we analyze 3 samples of quiescent galaxies using different diagnostics: (1) no or weak H$\beta$ emission line ($\mbox{EW(H}\beta\mbox{)} \geq -1$\AA), (2) low specific SFR (sSFR) based on the IR and UV luminosities ($sSFR_{UV+IR}< 10^{-10} M_\odot$yr$^{-1}$), and (3) the rest-frame U-V and V-J colors, the demarcation of \citet{muz13}. Table~\ref{tab:sample} lists the sample sizes with the 3 definitions of quiescence. The EW(H$\beta$) sample contains fewer galaxies because of an extra constraint on the spectral coverage; the spectra have to cover rest-frame $\sim 5000$\AA\ in order to measure the EW(H$\beta$), which is not required for the UV+IR sSFR and the UVJ color selections. The numbers in the parentheses are numbers of spectra which also covers the D$_n$4000 region. When D$_n$4000 is required for the analysis, we use this subset of data. 

These three classifications in general agree with each other well (Fig.~\ref{fig:sfq}). In each sample, $\sim90\%$ of galaxies are also defined as quiescent by the other two criteria. We estimate the `incompleteness' by calculating the number of star-forming galaxies that would be defined as quiescent galaxies by either of the other two criteria. The number of these potential missing objects is $<15\%$ of the number of quiescent galaxies. 

We adopt the the H$\beta$ emission line as our fiducial criteria of quiescence, as the Balmer emission lines trace star-formation in a shorter time-scale ($\sim10$~Myr) comparing to the other two indicators ($\gtrsim 100$~Myr). We note that we do not adopt the [O{\sc ii}]$\lambda$3727,3729 doublets as a star-formation indicator. 
It has been shown that roughly half of the red galaxies and PSB galaxies have moderate [O{\sc ii}] emission while lacking of Balmer emission. The excess of [O{\sc ii}] is likely due to AGN or LINERs rather than star-formation activities. Requiring no or weak [O{\sc ii}] emission results in a highly incomplete and potentially biased sample of quiescent galaxies \citep{yan06,lem10,wu14,lem17}. 

\begin{figure}
	\includegraphics[width=\columnwidth]{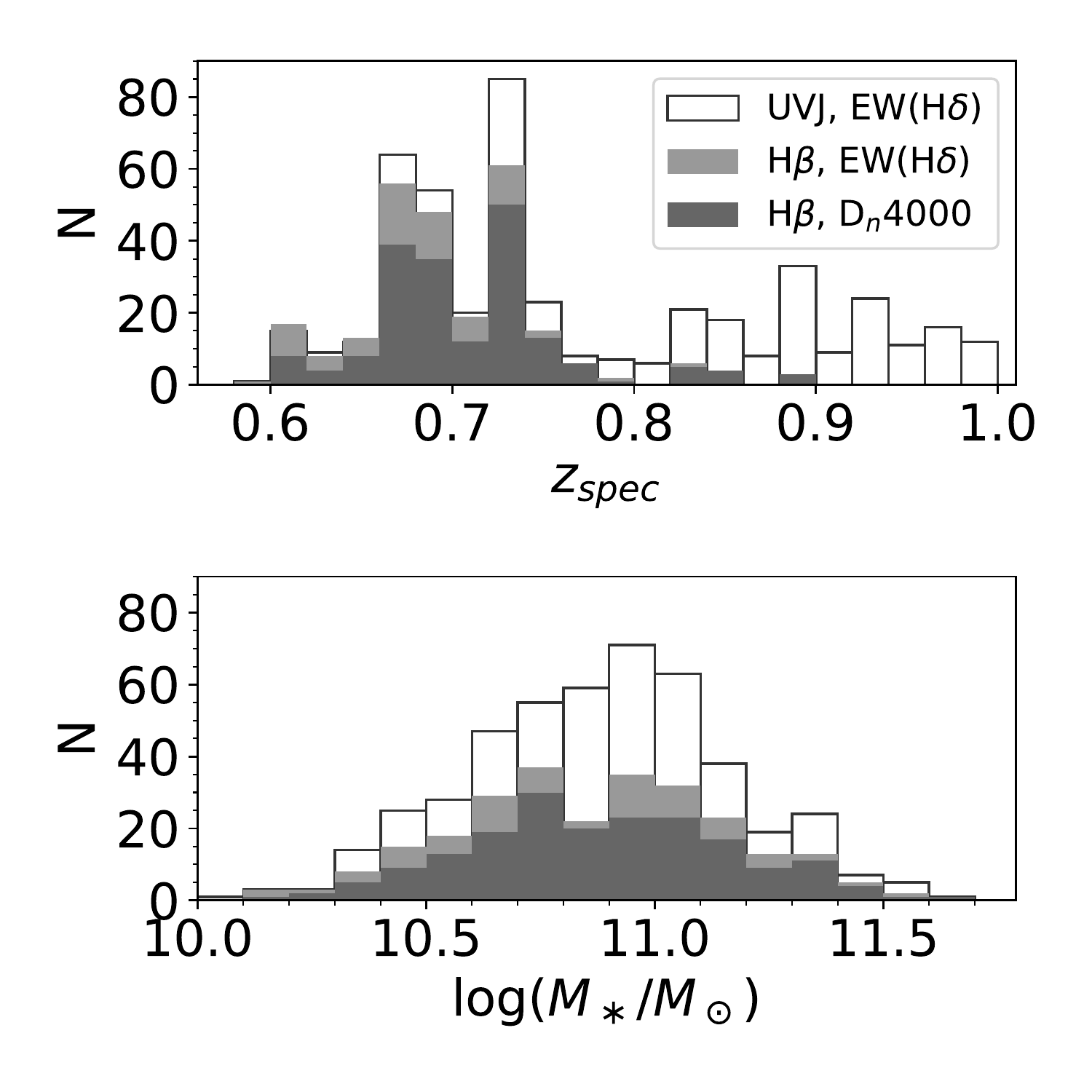}
	\caption{The distributions of redshifts and stellar masses of quiescent galaxies. The light and dark gray histograms are the distributions of galaxies with only EW(H$\delta$) and both EW(H$\delta$) and D$_n$4000 measurements, respectively. The distributions of UVJ quiescent sample is also presented for a comparison. The EW(H$\beta$) quiescent galaxies have $z\lesssim0.8$ because the H$\beta$ line shifts outside of our spectral coverage at $z\gtrsim0.8$; We are not able to use H$\beta$ to select quiescent galaxies. }
	\label{fig:mz}
\end{figure}

We repeat the analysis using the UV+IR sSFR and UVJ color quiescent samples and show the complementary results in the Appendix. The conclusions in this paper is not affected by the definition of quiescence. Fig.~\ref{fig:mz} shows the redshift and stellar mass distributions of the EW(H$\beta$) quiescent sample. We also plot the distributions of UVJ quiescent galaxies for a comparison. 
The EW(H$\beta$) quiescent galaxies are at lower redshifts because the H$\beta$ line is generally outside our spectral coverage at $z\gtrsim0.8$; we are not able to use the H$\beta$ selection for those galaxies.

\begin{table}
	\caption{Sample sizes}
	\label{tab:sample}
	\begin{center}
		\begin{threeparttable}
			\begin{tabular}{lrr}
				\hline
				\hline
				Quiescence Criteria  & $N$ & $N_{PSB}$\\
				\hline
				EW(H$\beta$)$> -1$\AA & 260(189) & 13(8) \\
				$sSFR_{UV+IR} < 10^{-10} M_\odot \mbox{yr}^{-1}$ & 441(372) & 15(13)  \\
				UVJ color & 467(390) & 35(31) \\
				\hline
			\end{tabular}
			\begin{tablenotes}
				\small
				\item The numbers in the parentheses are number of spectra covering both EW(H$\delta$) and D$_n$4000. 
			\end{tablenotes}
		\end{threeparttable}
	\end{center}
\end{table}

\subsection{Definition of Post-starburst galaxies}

PSB galaxies are galaxies whose star-formation was shut off rapidly in the last $\sim$1~Gyr. One of the prominent spectral features of PSB galaxies is their strong Balmer absorptions, indicating a dominating population of A to early F-type stars \citep{dre83,bal99,dre99}. 

In order to select PSB galaxies, several methods have been proposed for different types of data \citep[e.g.,][]{wil09,kri10,muz12}. Our spectra allow us to measure the H$\delta$ absorption strength accurately. Therefore, we use a traditional definition, define PSB galaxies as quiescent galaxies with strong H$\delta$ absorption. The boundary of ``strong" H$\delta$ is often drawn at $\mbox{EW}(\mbox{H}\delta) = 3\sim5$\AA. In this paper, we require PSB galaxies to have $\mbox{EW}(\mbox{H}\delta)>4$\AA. The numbers of PSB galaxies with each definition of quiescent is listed in Table~\ref{tab:sample}. 

\subsection{Mass size relation}
\label{sec:ms}
\begin{figure}
	\centering
	\includegraphics[width=\columnwidth]{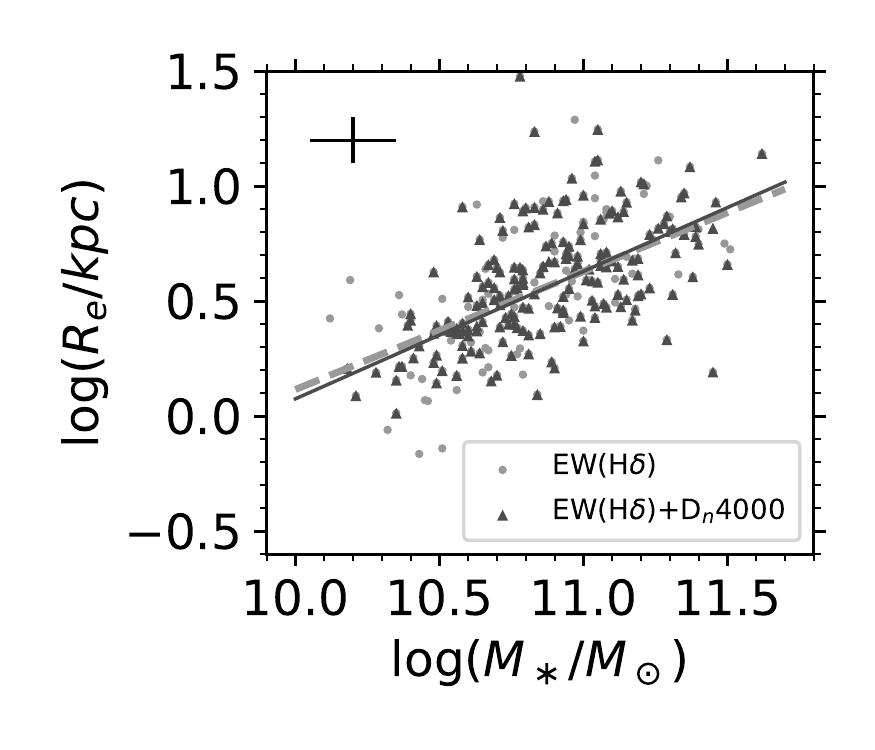}
	\caption{The sizes and stellar masses of quiescent galaxies. Dark gray triangles are galaxies whose spectra cover both D$_n$4000 and EW(H$\delta$). Light gray circles are galaxies for which only EW(H$\delta$) is measurable. The light gray dashed line and the dark gray solid line are the best-fit mass-size relation for all data and black points, respectively. The two best-fit relations are consistent with each other within 0.04~dex.}
	\label{fig:MR}
\end{figure}

Fig.~\ref{fig:MR} shows the mass-size relations of quiescent galaxies. 
We fit a linear relation between the median sizes in each 0.1~dex mass bins and the mass in the log-log space:
\begin{equation}
\label{eq:Re}
\log(R_e/kpc)_{med} = a \times [\log(M_\ast/M_\odot) - 11] + b.
\end{equation}
We fit both for the whole sample and the D$_n$4000 subset and estimate the uncertainties from bootstrap resampling. The two best-fit parameters $(a, b)$ are $(0.51^{+0.06}_{-0.06}, 0.63^{+0.01}_{-0.01})$ and $(0.55^{+0.04}_{-0.03}, 0.63^{+0.01}_{-0.01})$, respectively. The two best-fit relations agree with each other within 0.04~dex in $R_e$ at fixed mass. 

\begin{figure}
	\centering
	\includegraphics[width=\columnwidth]{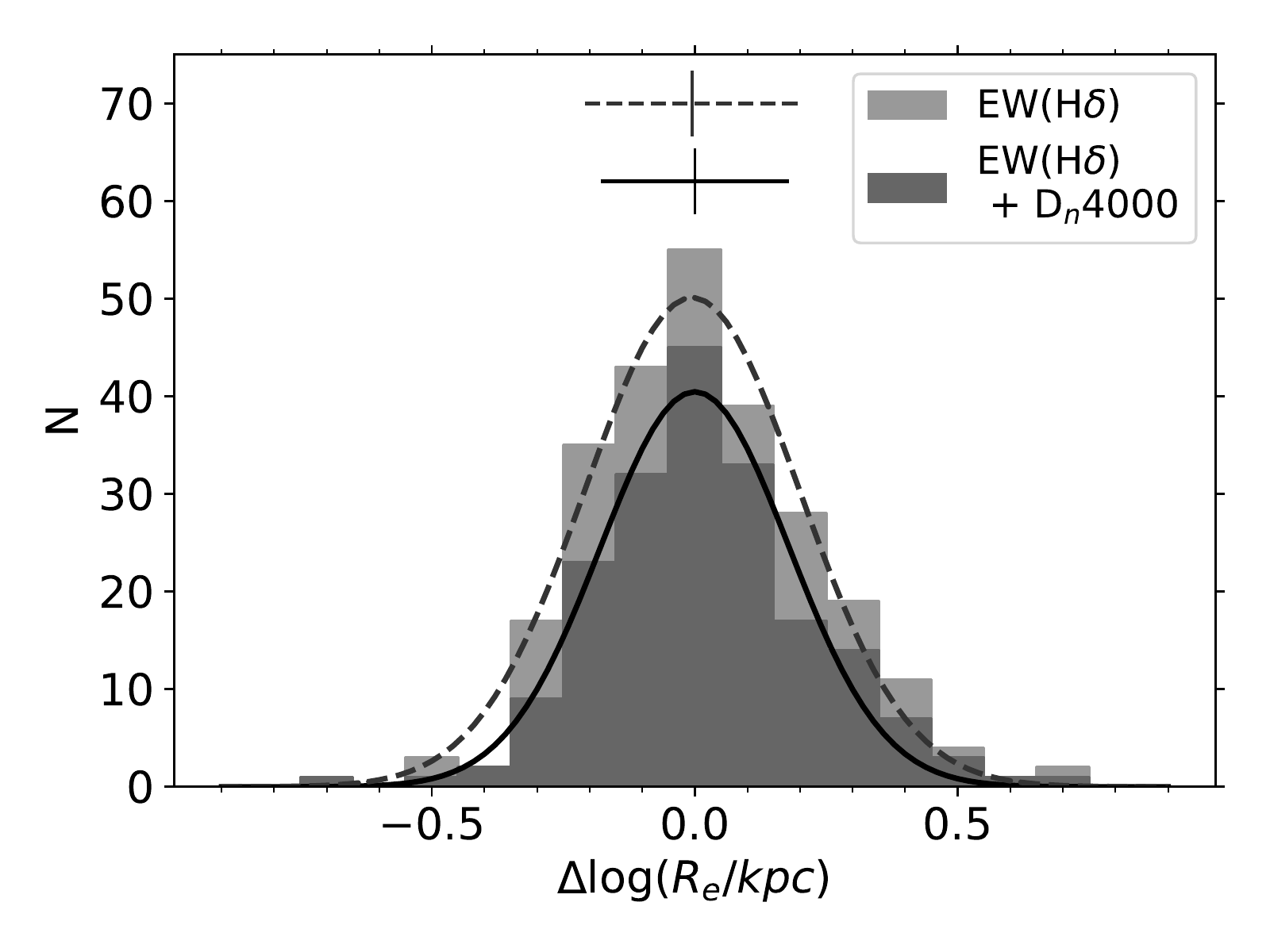}
	\caption{The distributions of $\Delta \log(R_e)$, the sizes relative to the median sizes at fixed stellar mass. The light gray histogram is the distribution of the whole quiescent sample. The dark gray histogram is the subset of galaxies which also have D$_n$4000 measurements. The dashed and the solid curves are best-fit Gaussian models to the two distributions. The distributions are well represented by Gaussian distributions centered at $\sim 0$ and with standard deviations of $\sim0.2$~dex as labeled at the top of the panel.} 
	\label{fig:drdist}
\end{figure}

In this paper, we use the sizes relative to the median sizes at fixed masses as the metric for galaxy sizes:
\begin{equation}
\Delta \log(R_e) = \log(R_e) - \log(R_e)_{med},
\end{equation}
where $\log(R_e)_{med}$ is the best-fit median size at given masses according to Equation~\ref{eq:Re} and the best-fit parameters of each sample, respectively. The distributions of $\Delta \log(R_e)$ can be modeled as a Gaussian distribution centered at $\sim0$, with a standard deviation of $\sim0.2$~dex (Fig.~\ref{fig:drdist}).

We note that our best-fit slope is different from the slope derived from the CANDLES survey \citep[0.71,][]{vdw14}. The difference can be a result of different sample selections and fitting methods.
We checked that the S/N cut does not bias the size. The $\Delta \log(R_e)$ distribution of low-S/N galaxies with good mass and size measurements can be well modeled by a Gaussian centered at $\sim0$ with a standard deviation of $\sim0.2$~dex, similar to Fig.~\ref{fig:drdist}. We also repeat the fitting using samples without the S/N cut and find that the slopes change by $<0.03$ and best-fit $R_e$ at fixed masses change by $<0.04$~dex. 

\section{Correlations between sizes and age-sensitive indicators}
\label{sec:res}

We examine the size dependences of two age-sensitive spectral features, D$_n$4000 and EW(H$\delta$). For passive galaxies, the D$_n$4000 increases monotonically with the stellar age. On the other hand, the EW(H$\delta$) is sensitive to recent star-formation such that galaxies with rapidly declining SFRs will have elevated EW(H$\delta$) in the first few hundred Myr after the star-formation stops then the EW(H$\delta$) gradually decreases afterward. These two indices together work as proxies for stellar ages and the recent star-formation histories \citep{kau03a,wu18b}.

\subsection{The dependence of D$_n$4000 and EW(H$\delta$) on galaxy size}
\label{sec:mr-age}
\begin{figure*}
	\centering
	\includegraphics[width=\textwidth]{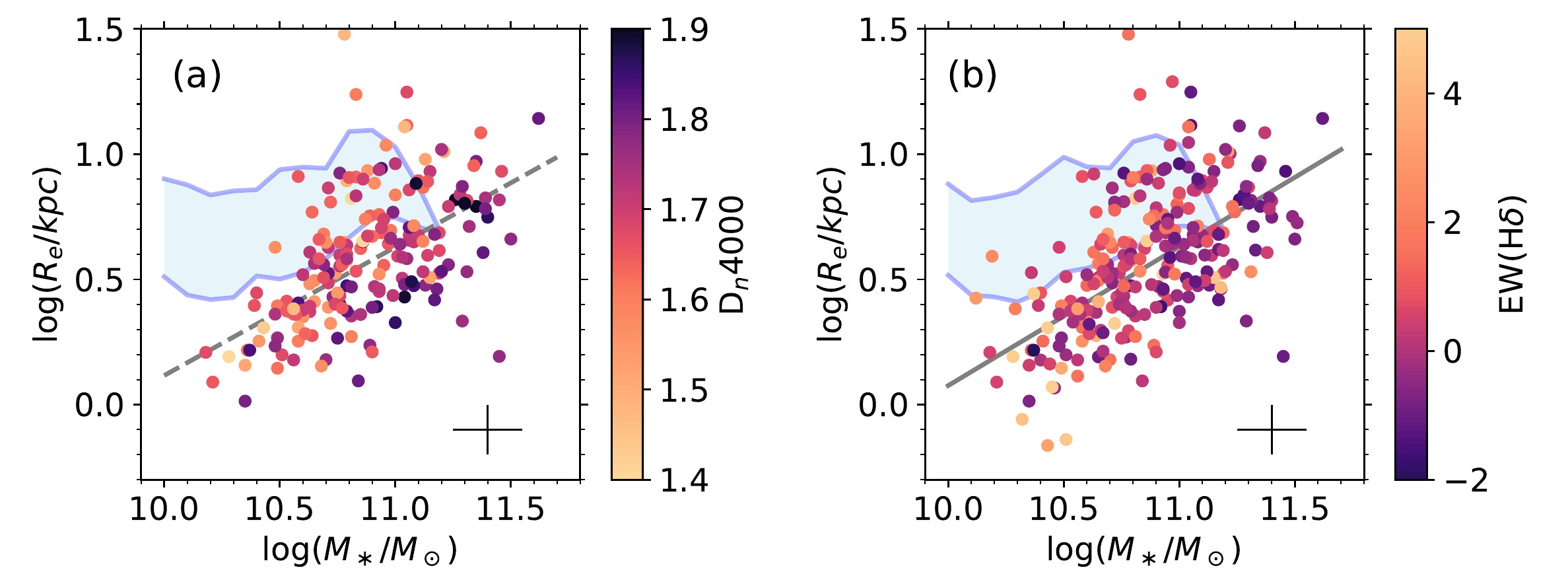}
	\caption{The masses and sizes of quiescent galaxies, color-coded by D$_n$4000 and EW(H$\delta$). At any fixed mass and size, galaxies can have a wide range of D$_n$4000 and EW(H$\delta$), indicating a large variation in stellar ages among galaxies of similar masses and sizes. There is a tentative trend such that more massive galaxies and smaller galaxies have larger D$_n$4000. The EW(H$\delta$) has a weak dependence on mass, where more massive galaxies have smaller EW(H$\delta$). The dependence on the size is not clear. The light-blue shaded areas are the 16th and 84th percentiles of the sizes of star-forming galaxies at fixed masses. The dashed and solid lines are the best-fit mass-size relations of quiescent galaxies. The cross in the bottom right corner represents the uncertainties in stellar mass (0.15~dex) and sizes (0.10~dex). }
	\label{fig:MRcolor}
\end{figure*}

Fig.~\ref{fig:MRcolor} shows the stellar masses and sizes of quiescent galaxies, color-coded by D$_n$4000 and EW(H$\delta$). For comparison, we also plot the size distribution of star-forming galaxies (16th and 84th percentiles) as a function of stellar masses. First of all, at a given mass and size, galaxies can have very different D$_n$4000 and EW(H$\delta$). The typical uncertainty of our D$_n$4000 and EW(H$\delta$) measurement is $\sim0.025$ and $\sim0.4$\AA, much smaller than the range of the distribution. 
The large color variations in Fig.~\ref{fig:MRcolor} are not due to measurement uncertainties but suggest large individual variations in ages at fixed stellar mass and size. 

Despite the large individual variation, a weak trend can be identified in Fig.~\ref{fig:MRcolor}a. Galaxies with large D$_n$4000, those with dark colors, tend to be massive and smaller than the average. The upper half of the mass-size relation is occupied mainly by points with lighter color, those galaxies with small D$_n$4000. As for the EW(H$\delta$) in Fig.~\ref{fig:MRcolor}b, a weak mass dependence is present such that high mass galaxies have on average smaller EW(H$\delta$). On the other hand, there is no clear size dependence by visual inspection. 

\begin{figure*}
	\centering
	\includegraphics[width=\textwidth]{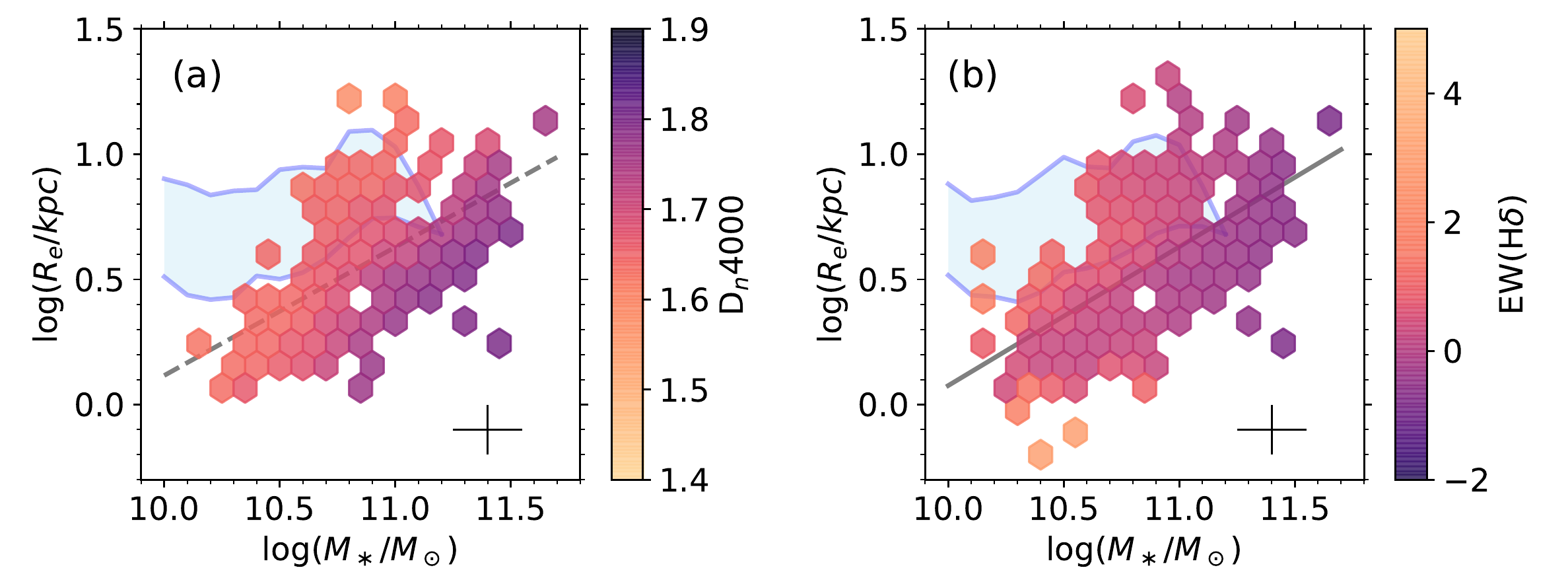}
	\caption{The D$_n$4000 and EW(H$\delta$) averaging over nearby data points on the mass-size plan using the LOESS method (see text). The smoothed maps reveal the underlying size dependence. Larger galaxies have on average smaller D$_n$4000 and larger EW(H$\delta$). However, the most compact galaxies also have large EW(H$\delta$), which is opposite to the general trend. The light-blue shaded areas are the 16th and 84th percentiles of the sizes of star-forming galaxies at fixed masses. The dashed and solid lines are the best-fit mass-size relations of quiescent galaxies. Dashed and solid lines are the best-fit mass-size relations of quiescent galaxies. Complementary results using other definition of quiescence are shown in Fig.~\ref{fig:MRloessUVIR} and Fig.~\ref{fig:MRloessUVJ} in the Appendix.}
	\label{fig:MRloess}
\end{figure*}

We apply the Locally Weighted Regression (LOESS) method to derive the mean trends of D$_n$4000 and EW(H$\delta$) on the mass-size plane (Fig.~\ref{fig:MRloess}). We use the \texttt{CAP\char`_LOESS\char`_2D} routine of \citet{cap13}, which implements the multivariate LOESS algorithm of \citet{cle88}. We adopt a regularization factor $f=0.5$ and a linear local approximation. 

\begin{figure*}
\centering
\includegraphics[width=\textwidth]{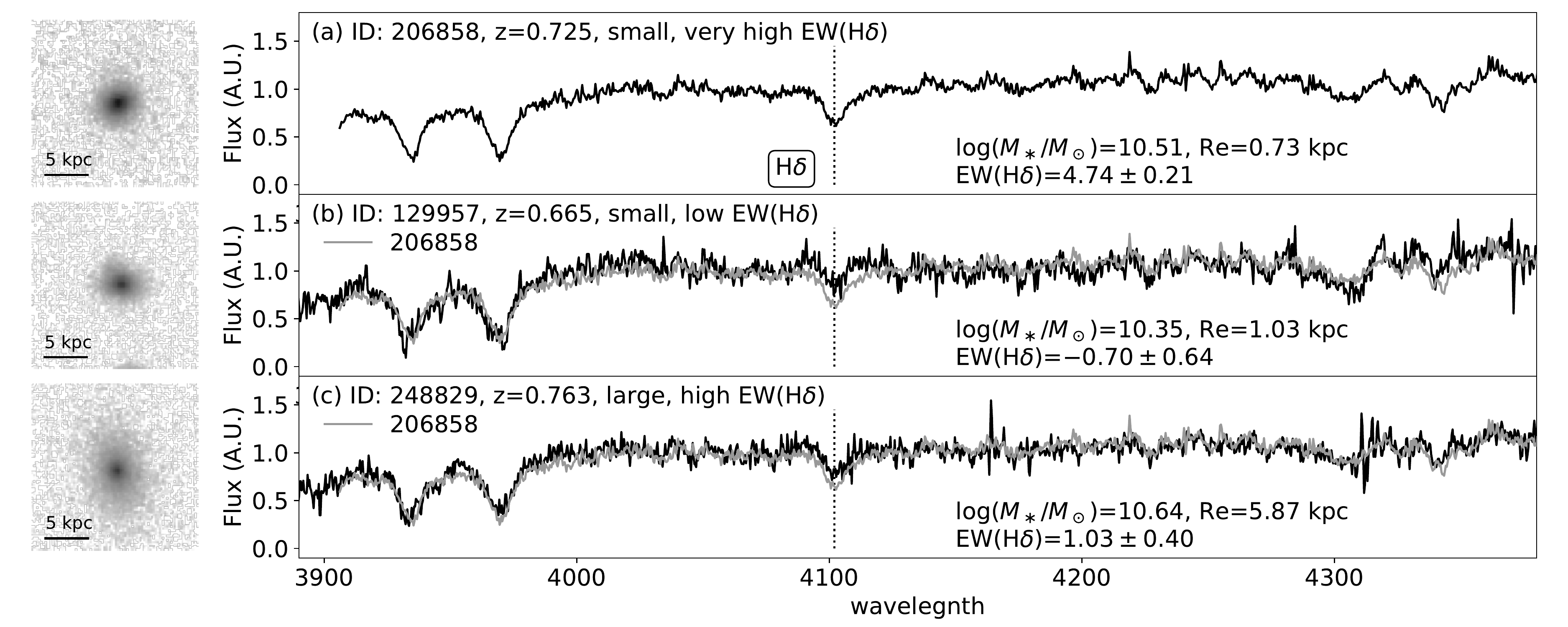}
\caption{Comparison of \textit{HST} F814W images and spectra of galaxies with similar stellar masses. All three galaxies have $\log(M_\ast/M_\odot)\simeq10.5$. (a) A small galaxy with very high EW(H$\delta$). (b) A small galaxy with low EW(H$\delta$). (c) A large galaxy with high EW(H$\delta$). The gray line in panel (b) and (c) are the spectrum of the galaxy in panel (a) for comparison. Our \textit{HST} images and spectra are able to differentiate among galaxies with different structural and spectral properties. The galaxy in panel (a), despite its small size, the strength of EW(H$\delta$) absorption is clearly stronger than the other two galaxies.}
\label{fig:rhd}
\end{figure*}

The LOESS-smoothed maps confirm the visual inspection on Fig.~\ref{fig:MRcolor}. Firstly, more massive galaxies have on average larger D$_n$4000 and smaller EW(H$\delta$), suggesting that massive galaxies form their stars earlier \citep[see][for results at similar redshifts]{hai17,siu17,wu18b}. At fixed stellar mass, larger galaxies have smaller D$_n$4000 \citep[see][for results at lower redshifts]{zah17}.
On the other hand, the behavior of EW(H$\delta$) is more complex. For massive galaxies ($M_\odot \gtrsim 10^{11}M_\odot$), the LOESS-smoothed map does not show a clear trend in EW(H$\delta$) as a function of the size. For the bulk of the lower mass galaxies, larger galaxies have slightly larger EW(H$\delta$). 
However, we also find that the smallest bins have large EW(H$\delta$) (see also Fig.~\ref{fig:MRloessUVIR} and Fig.~\ref{fig:MRloessUVJ} in the Appendix). This reversal is driven by a few small galaxies with very strong H$\delta$ absorption (Fig.~\ref{fig:MRcolor}). Fig.~\ref{fig:rhd}a shows an example. Comparing to galaxies of similar masses, the Balmer absorption is clearly stronger than the galaxy of similar size (Fig.~\ref{fig:rhd}b) or a much larger galaxy (Fig.~\ref{fig:rhd}c). We discuss the correlation between the spectral indices and the size of galaxies in further detail in the following sections. 

\subsection{The correlation between stellar ages and sizes of quiescent galaxies}
\label{sec:AR}

\begin{figure*}
	\centering
	\includegraphics[width=0.9\textwidth]{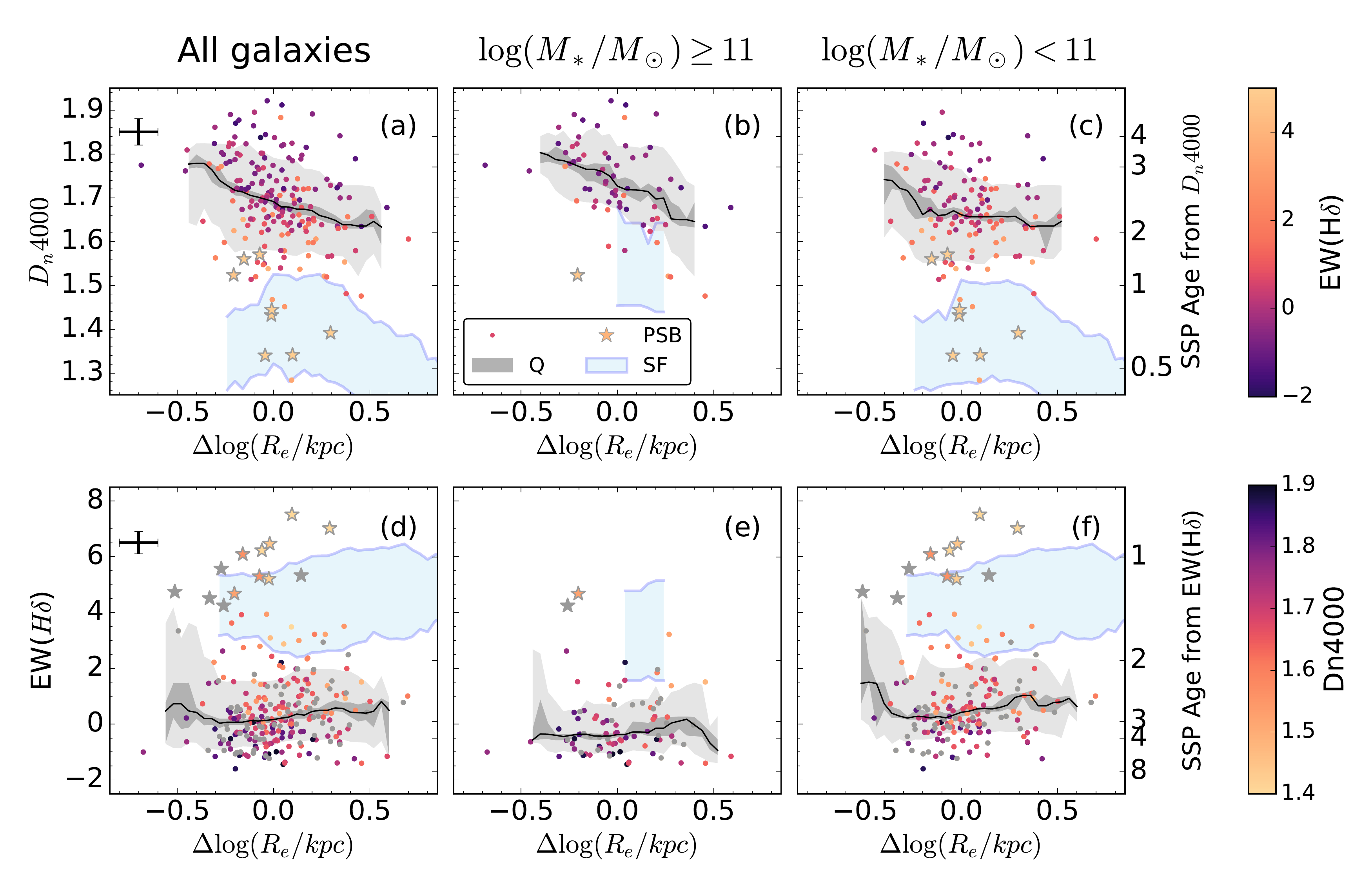}
	\caption{The correlation between the size and the absorption line indices D$_n$4000 and EW(H$\delta$) of all (left), high-mass (center), and low-mass (right) quiescent galaxies. PSB galaxies are labeled as stars. In panels (a), (b), and (c), the color is the EW(H$\delta$) value. In panels (d), (e), and (f), the color is the D$_n$4000. Gray points are galaxies without D$_n$4000 measurements. The solid black lines show the running median of D$_n$4000 and EW(H$\delta$) as a function of $\Delta\log(R_e)$, the sizes relative to the median sizes at fixed masses. The light gray areas indicate the 16th and 84th percentiles of the distributions at fixed $\Delta\log(R_e)$. The dark gray areas are the uncertainties of the medians, estimated from 1000 bootstrap samples. The light blue area shows the 16th and 84th percentiles of the indices of star-forming galaxies. The y-axis on the right hand side labels the corresponding ages of a solar metallicity SSP of each indice. Star-forming galaxies have on average smaller D$_n$4000, larger EW(H$\delta$), and larger sizes than quiescent galaxies. We only show interval with more than 5 galaxies within a 0.2~dex size bin. Overall, the D$_n$4000 of quiescent galaxies decreases as $\Delta\log(R_e)$ increases, indicating that larger galaxies are on average younger. There is no clear size dependence on EW(H$\delta$). At fixed $\Delta\log(R_e)$, the distributions of D$_n$4000 and EW(H$\delta$) correspond to an age range of $\sim1.5$ Gyr assuming a solar metallicity SSP.}
	\label{fig:AR}
\end{figure*}

Fig.~\ref{fig:AR} shows the relations between $\Delta\log (R_e)$, the size relative to the median sizes at fixed mass (Section~\ref{sec:ms}), and D$_n$4000 and EW(H$\delta$) for the whole sample and in two stellar mass bins. As in Fig.~\ref{fig:MRcolor}, at fixed $\Delta \log R_e$, quiescent galaxies can have very different D$_n$4000 and EW(H$\delta$). The light gray shaded areas in Fig.~\ref{fig:AR}, the 16th and the 84th percentiles of the D$_n$4000 and EW(H$\delta$) distributions, cover $\mbox{D}_n4000 \simeq 1.6-1.8$ and $\mbox{EW}(\mbox{H}\delta) \simeq -1\AA - 2\AA$, respectively. For a simple stellar population SSP with solar metallicity \citep{bc03}, these ranges correspond to an age range of $\sim1.5$~Gyr. The running medians (black lines) show that the D$_n$4000 progressively decreases as $\Delta\log (R_e)$ increases. The overall size dependence in EW(H$\delta$) is less significant. For the low-mass bin, larger galaxies have on average slightly larger EW(H$\delta$), except for the most compact galaxies. There is no detectable size dependence for the high-mass bin.

\begin{figure*}
	\includegraphics[width=\textwidth]{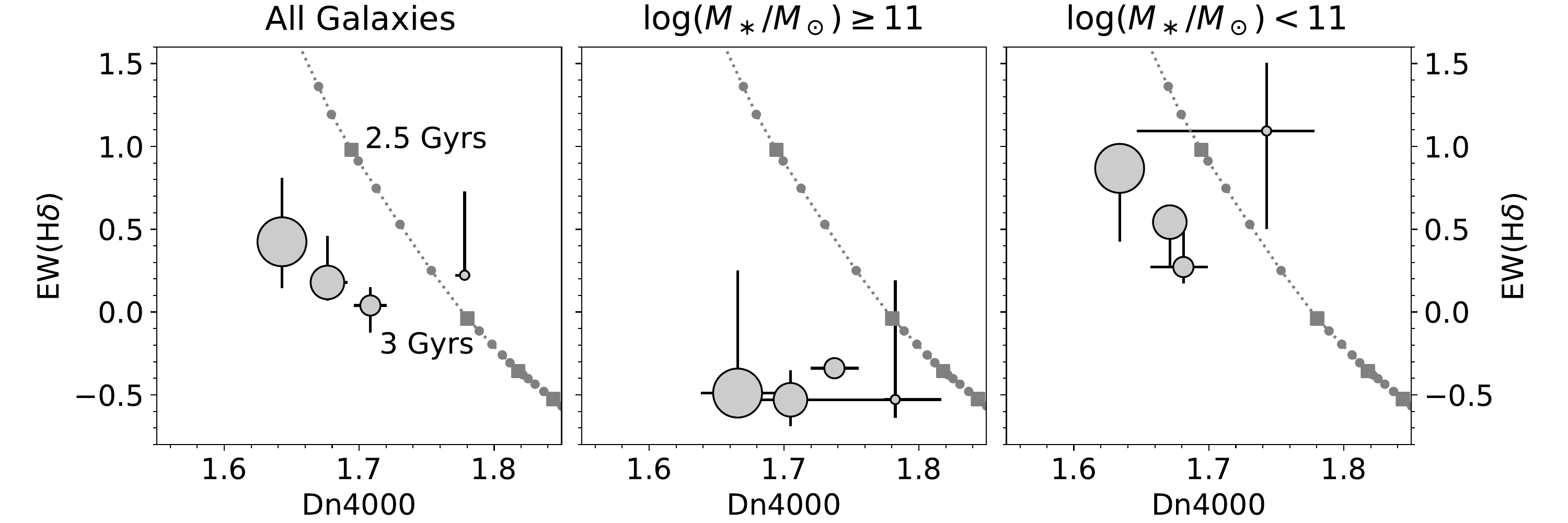}
	\caption{The D$_n$4000 and EW(H$\delta$) of quiescent galaxies in 4 size bins: $\Delta \log(R_e) > 0.2$, $0.2 \geq \Delta \log(R_e) > 0$, $0 \geq \Delta \log(R_e) > -0.2$, and $-0.2 \geq \Delta \log(R_e)$. All, high-mass, and low-mass galaxies are shown separately. In each panel, four circles represent the median D$_n$4000 and EW(H$\delta$) in each size bin, where larger circles correspond to larger galaxies. The uncertainties of medians are calculated from 1000 bootstrap samples. The dotted lines are galaxy evolutionary model tracks for an SSP with solar metallicity. Squares mark the age of 2.5, 3.0, 3.5, and 4.0 Gyrs. Small gray dots label each of 0.1~Gyr time stamp. The 3 larger bins show the correlation between the ages and the sizes; larger galaxies are on average younger. Nevertheless, this correlation is less significant for massive galaxies.}
	\label{fig:ARDnHd}
\end{figure*}

We show the median indices of galaxies in 4 size bins on the D$_n$4000-EW(H$\delta$) plane (Fig.~\ref{fig:ARDnHd}). The size bins are divided at $\Delta \log(R_e) =$ -0.2, 0, and 0.2. The middle two size bins contain the majority of galaxies as 0.2~dex is the $1\sigma$ of the distribution of $\Delta \log (R_e)$ (Section~\ref{sec:ms}). The largest and the smallest bins, on the other hand, contain galaxies with size significantly larger or smaller than the average population. We also plot the evolutionary model track of the solar-metallicity SSP for a reference. The trend in D$_n$4000 and EW(H$\delta$) combined suggests that except for the smallest bin, there is a weak size dependence on stellar ages that larger galaxies are on average younger, but the difference in age is small, $<500$~Myrs. Galaxies in the smallest size bin are a mix of old galaxies and post-starburst galaxies (See Section~\ref{sec:PSB}).

\begin{figure*}
	\centering
	\includegraphics[width=0.9\textwidth]{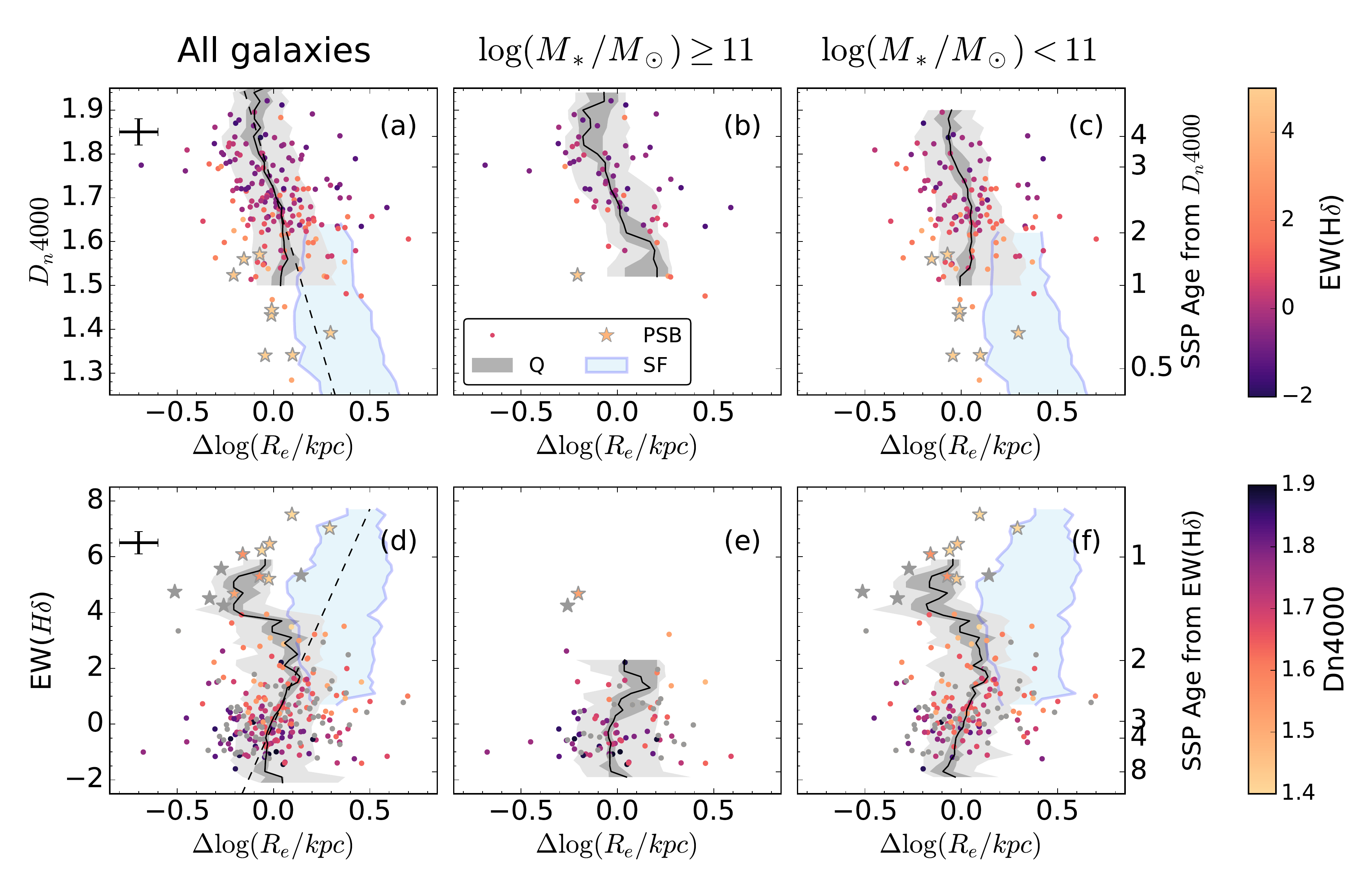}
	\caption{The sizes as a function of D$_n$4000 and EW(H$\delta$) of all (left), high-mass (center), and low-mass (right) quiescent galaxies. PSB galaxies are labeled as stars. In panels (a), (b), and (c), the color is the EW(H$\delta$) value. In panels (d), (e), and (f), the color is the D$_n$4000, and galaxies without D$_n$4000 measurements are in gray. The data points are the same as in Fig.~\ref{fig:AR}. The solid black lines show the running median of $\Delta\log(R_e)$, at fixed D$_n$4000 or EW(H$\delta$). The light gray areas indicate the 16th and 84th percentiles of the distribution at fixed indices. The dark gray areas are the uncertainties of median $\Delta \log(R_e)$, estimated from 1000 bootstrap samples. The light blue area shows the 16th and the 84th percentiles of $\Delta \log(R_e)$ of star-forming galaxies for a comparison. The y-aixs on the right hand side labels the corresponding ages of a solar metallicity SSP of each indice. We only show intervals with more than 5 galaxies in bins of 0.05 in D$_n$4000 and 0.8\AA\ in EW(H$\delta$). There are too few massive star-forming galaxies for plotting. The dashed lines in panels (a) and (d) show the best-fit relations between $\Delta\log(R_e)$ and indices at $\mbox{D}_n4000 \geq 1.6$ and $\mbox{EW}(\mbox{H}\delta) \leq 2$\AA\ (Equation~\ref{eq:rind}) and the extrapolation beyond the fitting ranges. The quiescent and star-forming galaxies form a continuous sequence on the D$_n$4000--$\Delta \log(R_e)$ plane. As D$_n$4000 decreases, the median size increases. The EW(H$\delta$) shows a qualitative same behavior except for galaxies with the highest EW(H$\delta$). PSB galaxies are much smaller than star-forming galaxies with the same EW(H$\delta$).}
	\label{fig:RA}
\end{figure*}

We also investigate how the sizes of galaxies change as a function of D$_n$4000 or EW(H$\delta$) (Fig.~\ref{fig:RA}). The $\Delta \log(R_e)$ increases towards smaller D$_n$4000 and larger EW(H$\delta$) at $\mbox{D}_n4000 \gtrsim 1.6$ and $\mbox{EW}(\mbox{H}\delta) \lesssim 2$\AA, the regimes where the quiescent galaxies are the dominating population (Fig.~\ref{fig:AR} and \citet{kau03a,hai17,wu18b}).  
Within these ranges of indices, the Spearman's rank correlation coefficients $\rho$ and their p-values between $\Delta \log(R_e)$ and indices are $\rho=-0.37$, $\mbox{p-value}=3\times10^{-6}$ for D$_n$4000 and $\rho=0.19$, $\mbox{p-value}=4\times10^{-3}$ for EW(H$\delta$).
We fit linear relations to the median $\Delta \log(R_e)$ and the indices
for quiescent galaxies with $\mbox{D}_n4000 \geq 1.6$ and $\mbox{EW}(\mbox{H}\delta) \leq 2$\AA\ in Fig.~\ref{fig:RA}a and Fig.~\ref{fig:RA}d as:
\begin{equation}
\begin{aligned}
\Delta \log(R_e) = [(-0.68\pm0.08) \times \mbox{D}_n4000] + (1.17\pm0.14) \\ \Delta \log(R_e) = [(0.065\pm0.005) \times \mbox{EW}(\mbox{H}\delta)] + (-0.001\pm0.005)
\end{aligned}
\label{eq:rind}
\end{equation}
These results suggest a trend that quiescent galaxies with younger stellar populations are on average larger, consistent with Fig.~\ref{fig:AR} and Fig.~\ref{fig:ARDnHd}. 

However, the median sizes become smaller at $\mbox{EW}(\mbox{H}\delta) > 2$\AA, reaching $\Delta \log(R_e) \simeq -0.2$ at $\mbox{EW}(\mbox{H}\delta) \simeq4$\AA\ (Fig.~\ref{fig:RA}d and Fig.~\ref{fig:RA}f). We will discuss the sizes of these PSB galaxies in the next section.

\subsection{The size of Post-starburst Galaxies}
\label{sec:PSB}

\begin{figure}
	\includegraphics[width=\columnwidth]{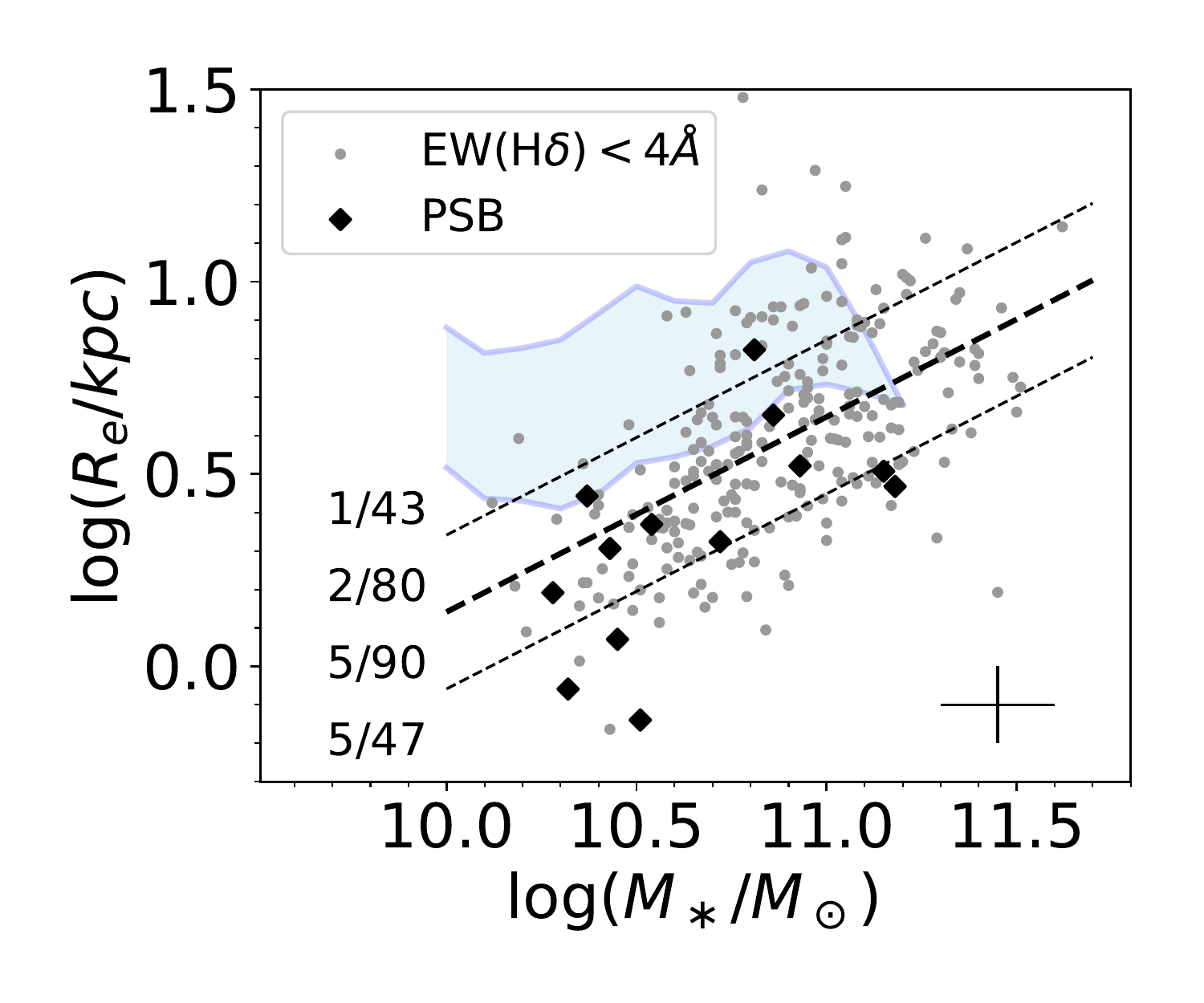}
	\caption{The distribution of PSB galaxies on the mass-size plane. Big black diamonds are quiescent galaxies with $\mbox{EW(H}\delta\mbox{)} \geq 4$\AA, defined as PSB galaxies. Other quiescent galaxies are the small gray dots. The thick dashed line is the best fit mass-size relation. The thin dashed lines label the $\pm0.2$~dex ($\sim1\sigma$) range around the best-fit relation. The blue shaded area shows the 16th and 84th percentiles of the sizes of star-forming galaxies at fixed stellar mass. The numbers of all quiescent galaxies and PSB galaxies in 4 different ranges of sizes are labeled at the left. The majority of PSB galaxies (10/13) are smaller than the average size of quiescent at fixed stellar mass. About 40\% of PSB galaxies (5/13) are located below the $1\sigma$ distribution. Furthermore, most PSB galaxies are significantly smaller than star-forming galaxies. }
	\label{fig:PSB}
\end{figure}

Fig.~\ref{fig:PSB} shows PSB galaxies on the mass-size plane. The majority (10/13) of PSB galaxies are located at the bottom half of the mass-size plane and $\sim$40\% (5/13) are below 1$\sigma$ of the size distribution (see Fig.~\ref{fig:PSBapp} in the Appendix for complementary results). The PSB galaxies have median $\mbox{D}_n4000 \simeq1.45$ and median $\mbox{EW}(\mbox{H}\delta)\simeq5.2$\AA. Casting the indices in terms of the ages of a SSP, they are among the youngest quiescent galaxies but they are not larger than the average quiescent population.

The whole PSB population has a median size of $\Delta \log(R_e) = -0.07\pm0.07$. By extrapolating Equ.~\ref{eq:rind}, the expected sizes of galaxies with $\mbox{EW}(\mbox{H}\delta)\simeq5$\AA\ will be $\Delta \log(R_e) \simeq 0.3$, which is much larger than PSB galaxies but comparable to star-forming galaxies (See Fig.~\ref{fig:RA}d). 
Even comparing to the sizes of typical young quiescent galaxies ($\Delta \log(R_e) \simeq 0.13$ at $\mbox{EW}(\mbox{H}\delta)\simeq2$\AA), PSB galaxies are still significantly smaller; They do not follow the general correlation between the age and the size characterized by the majority of galaxies.

\section{Discussion}
\label{sec:dis}

The results in Section~\ref{sec:res} can be summarized as at fixed mass, (1) larger quiescent galaxies have on average younger stellar ages, and (2) PSB galaxies are much smaller than expected based on their young stellar populations.
The correlation among the sizes, the ages, and the recent star-formation histories puts constraints on the formation and evolution processes of quiescent galaxies. Several recent studies address this question but report seemingly contradictory results. In this section, we show that our high-quality data and the larger sample size present a complete picture. We then discuss the implication on the formation of quiescent galaxies.

\subsection{Reconciling tensions in the literature}

Several studies have investigated the age-size correlation of quiescent galaxies. \citet{fag16} and \cite{wil17} split their samples into two size bins and derive galaxy ages from co-added spectra in each size bin. They find that at $M_\ast < 10^{11} M_\odot$, smaller galaxies are slightly older, but the age difference between galaxies in the two size bins is at largest a few hundred Myrs at $0.2 < z < 1.2$. For galaxies with $M_\ast > 10^{11} M_\odot$, there appears to be no clear trend. \citet{bel15} shows that at $1<z<1.5$, quiescent galaxies older than 1.25~Gyr are smaller than the average at fixed stellar masses, whereas \citet{tru11} and \citet{zan16} find no significant size difference between old and young quiescent galaxies at $z<2$. 

On the other hand, \citet{kea15} find that compact massive galaxies ($r<2$~kpc and $M_\ast > 10^{11} M_\odot$) at $z\sim1$ are younger than a control sample of galaxies of similar masses. A few studies on high-$z$ PSB galaxies also find that these galaxies are among the smallest quiescent galaxies at fixed stellar masses \citep{whi12,yan16,alm17}. These studies based on either the most compact galaxies or PSB galaxies conclude that quiescent galaxies were compact when they formed. 

There is apparent tension between these statements. Our results in Section~\ref{sec:res} present a complete picture that reconciles these seemingly contradictory observations.
First of all, for the 3 largest size bins, the median D$_n$4000 and EW(H$\delta$) differ by $\sim0.06$ and $\sim0.5$\AA. When casting these age indicators in terms of single stellar population ages, the ages differ by only $<500$~Myr (Fig.~\ref{fig:ARDnHd}). The difference in age is in broad agreement with previous studies derived from stacking spectra of large and small galaxies \citep{tru11,fag16,wil17}. At the same time, we also find that PSB galaxies are much smaller than expected based on the global correlation between age and size.

The apparent contradiction in the literature is mainly due to sample selection and data binning that present only a partial view. 
First of all, studies focusing on special objects, either compact galaxies or PSB galaxies, pick up outliers that do not follow the average age-size correlation (Fig.~\ref{fig:RA}). Secondly, the individual age variation is large comparing to the average trend (Fig.~\ref{fig:AR}). The correlation between the age indicators and the sizes may not be detectable using a small sample or low S/N data. The stacking analysis can improve the measurements and recover the average trend when a large sample is available, but wash out the variations among individual galaxies. PSB galaxies represent only $\sim 5\%$ of the quiescent population in our sample \citep[13 out of 260 in our sample, see also][for studies at $z\sim1$]{ver08,yan09,muz12,wu14,wil16}. The stacking analysis cannot identify the peculiar behaviors of these rare special objects. Only with LEGA-C's unique combination of sample size and data quality, we can reveal the complex correlations between galaxy sizes and age indicators. 

\subsection{Multiple ways to quiescence}
The different sizes of PSB galaxies and the rest of young quiescent galaxies suggest that these populations have different star-formation histories. By definition, the star-formation activities in PSB galaxies have been shut off rapidly within a short timescale of a few hundred Myrs \citep{leb06,wil09,sny11}. On the contrary, other young quiescent galaxies did not necessarily experience the same rapid quenching processes. The correlation between the size and the timescale that star-formation shuts off corresponds to different types of evolutionary scenarios. 

\subsubsection{Slow process, large young quiescent galaxies}

The first type of mechanisms to stop star-formation is cutting off the supply of cold gas, therefore, galaxies naturally run out of fuel to form new stars. In massive dark matter halos, cold gas can be shock heated to high temperature while falling into the halos \citep{ker05,dek09}. Furthermore, active galactic nuclei (AGN) can inject energy and keep the halo gas hot \citep{bes05,cro06,sij07,bes14,yes14,ter16,bar17}. The star-formation rates thus gradually drop as the cold gas reservoir depletes. This process does not directly require a change in structure when a galaxy transforms from star-forming to quiescent. 

At fixed stellar mass, the size distributions of star-forming and quiescent galaxies overlap with each other in such that the smaller half of star-forming galaxies have similar sizes to the larger half of quiescent galaxies \citep[][also Fig.~\ref{fig:PSB}]{vdw14}. The similarity in stellar masses and sizes makes small star-forming galaxies candidate immediate progenitors of quiescent galaxies. As the SFR declines, a relative small star-forming galaxy naturally becomes a relative large quiescent galaxy with young stellar populations. The average size of the quiescent population also increases. 

\citet{car13} calculated the average size growth of the quiescent population at $z<1$ under the assumption that at any time, newly-formed quiescent galaxies have the same size distributions as the star-forming galaxies at the same epoch. They are able to reproduce the size evolution of massive ($M_\ast > 10^{11} M_\odot$) quiescent galaxies but over-predict the sizes of lower mass quiescent galaxies. If small star-forming galaxies are progenitors of newly-formed quiescent galaxies, this discrepancy can be mitigated.

In the local Universe, where the total cold gas mass is measurable, the gas content in star-forming galaxies strongly correlates with the size of stellar disks, such that at fixed stellar mass, star-forming galaxies with more compact disks possess less gas \citep{wu18a}. This dependence of gas content on disk sizes is likely in place up to $z\sim3$ based on an inversion of a star formation law \citep{pop15}. Once there is no more cold gas supply, smaller star-forming galaxies would use up the fuel quicker.
 
Moreover, from a dynamical point of view, more compact, high surface density galaxies are expected to be less stable against both global and local growth of instability thus more efficient in turning gas into stars \citep{dal97,mih97}. This size dependence of star-formation efficiency is also required to explain the size dependence of gas-phase metallicity \citep{wu15}. The instability may also trigger the gas flows toward the centers of galaxies, which accelerate the growth of the black holes \citep{bou11,gab11} and enhance the AGN feedback to keep the galaxies quiescent at earlier times. The cosmological simulation IllustrisTNG \citep{pil18} also suggests that smaller star-forming galaxies terminate their star-formation activities earlier \citep{gen18}.

\subsubsection{Fast process, PSB galaxies with various sizes}

The other type of mechanisms involve more violent events. Galaxy mergers, interaction, or violent disk instability can efficiently funnel gas into the centers of galaxies, inducing intense star formation in the galaxy centers and exhausting available gas in a short period of time \citep{yan08,dek09,zol15}. In addition, AGN triggered by infalling gas and strong supernova feedback after the starburst could eject ISM from galaxies, accelerates the quenching process \citep{efs00,spr05,hop06,hop08}. The merger can also enhance turbulence and make the gas stable against gravitational collapse \citep{ell18}. The rapidly rising star-formation rate and the subsequent quenching process can happen within a few hundreds of Myrs.
In this scenario, the last period of star-formation can build up a dense core in the centers. The descendant galaxy can thus have smaller $R_e$ than its progenitor \citep{bar91,mih94,bar96,bou11,dek14,zol15}.

Recent works have identified candidate progenitors of compact PSB galaxies. Case studies on some compact star-forming galaxies at $z\sim2$ found that they have centrally-concentrated cores in rest-frame UV and optical wavelengths and short gas depletion time \citep{bar16,barr17,pop17}, showing that these galaxies are possibly building the central dense cores and rapidly exhausting the gas reservoir simultaneously. After star-formation stops, the stellar population is dominated by A-type stars and the quiescent descendants would be classified as compact PSB galaxies.

\subsubsection{Coexistence of Multiple Pathways to Quiescence}

\begin{figure*}
	\centering
	\includegraphics[width=0.8\textwidth]{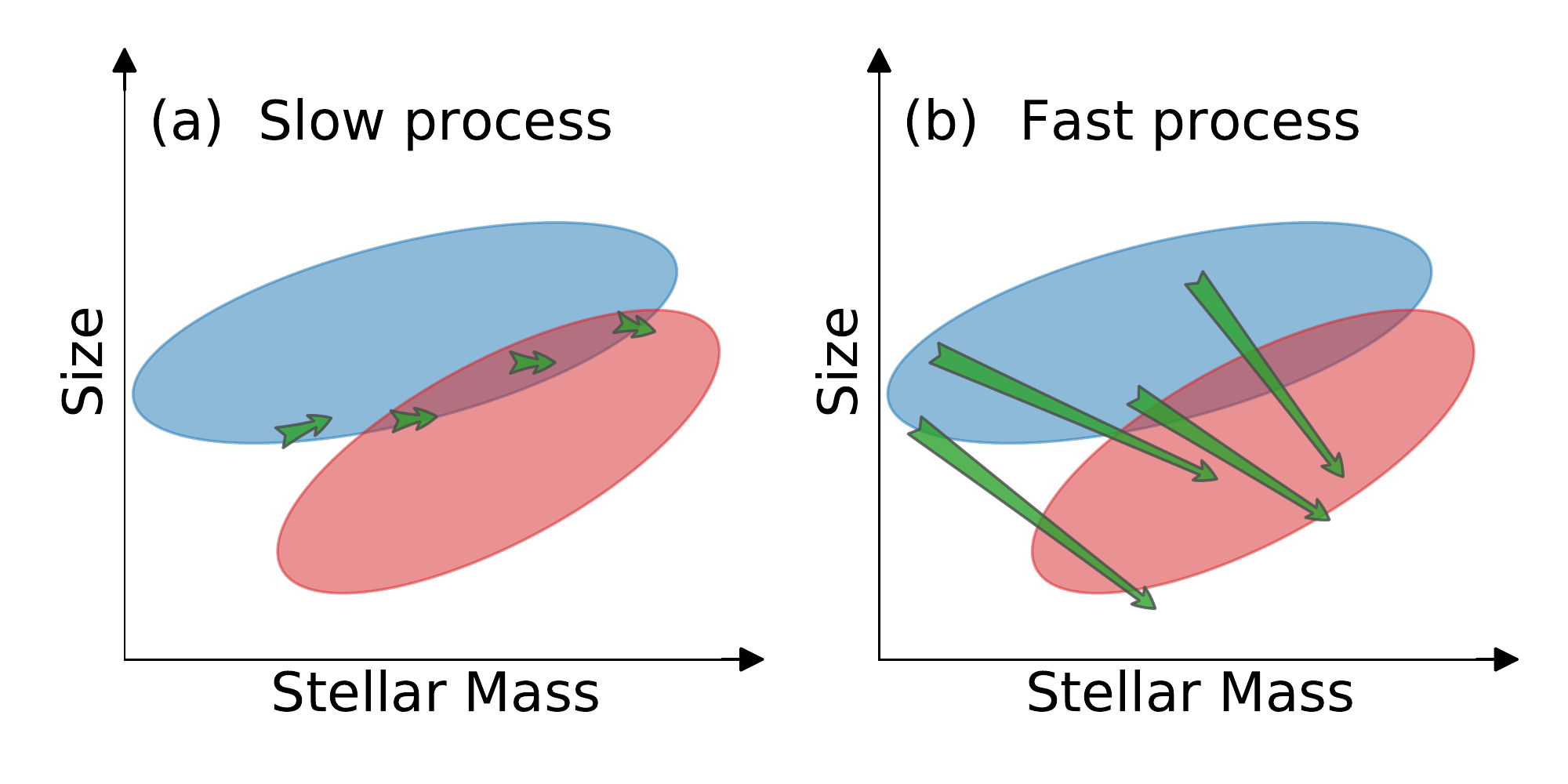}
	\caption{The evolution of the slow and the fast process on the mass-size plan. The blue and red shaded areas represent star-forming and quiescent populations, respectively. The arrows show how galaxies evolve on the mass-size plane relative to the bulk of the population when star-forming galaxies start to leave the main sequence to the time they are defined as quiescent. (a) Galaxies going through the slow process do not significantly change their masses and sizes. The stellar masses become slightly larger as the residual cold gas turns into stars. Although galaxy grow, the sizes measured in optical light may become slightly smaller after the extended star-forming disks fade. New quiescent galaxies are thus larger than the bulk of the quiescent population. (b) Galaxies going through the fast process can change their masses and sizes significantly. The masses increase due to episodes of intense star-formation. The sizes, quantified as $R_e$, decrease because the newly-formed stars are centrally concentrated. Depending on the masses and the spatial distribution of newly-formed stars, new quiescent galaxies can be smaller than existing quiescent population.}
	\label{fig:sch}
\end{figure*}

Our results suggest strongly that there are at least two types of mechanisms with distinct timescales in effect. Neither of the processes discussed above alone can easily produce the complex relationships among the size, the ages, and the recent star-formation history. 

First of all, simply cutting off the supply of cold gas cannot produce PSB galaxies. Comparisons of the cold gas content and the SFRs of star-forming galaxies at $z<1$ generally show that the gas depletion timescale is of the order of 1~Gyr or longer \citep{ler08,sch10,bie11,sai11,sch11,hua12,tac13,gen15,tac18}. Galaxies with such star-formation histories unlikely go through the post-starburst phase. When the SFR becomes low enough that galaxies are classified as quiescent, the stellar populations have already aged and do not exhibit strong Balmer absorption \citep{leb06}. To produce PSB galaxies, a much shorter quenching timescale ($\lesssim$ a few hundred Myr) or an additional starburst is necessary \citep{leb06,wil09,sny11}. Moreover, the size of PSB galaxies are significantly smaller than star-forming galaxies of the same mass (Fig.~\ref{fig:RA} and Fig.~\ref{fig:PSB}). To transform a star-forming galaxy to a PSB galaxy, both the SFR and the size must be altered by the quenching mechanism.

Secondly, PSB galaxies are unlikely the progenitors of the majority of quiescent population, at least up to $z\sim1$. If the size distribution of PSB galaxies is representative for all new quiescent galaxies, there would not be the age-size correlation that large galaxies are typically younger than smaller galaxies.

The complex correlation among sizes, ages, and recent star-formation histories warrants multiple mechanisms transforming galaxies from star-forming to quiescent \citep[also see ][]{sch14,bel15,mal18}. Fig.~\ref{fig:sch} illustrates these two different processes on the mass-size plane. 
The red and blue shaded areas represent star-forming and quiescent populations, respectively. The arrows show how galaxies evolve on the mass-size plane relative to the bulk of the population when star-forming galaxies start to leave the main sequence to the time they are defined as quiescent. 

Galaxies going through the slow process do not change their stellar masses and sizes much along with the SFRs (Fig.~\ref{fig:sch}a). Galaxies can grow slightly in mass and size due to the residual star formation \citep{tac15,tacc18}. Meanwhile, after the star-forming disk fades, the $R_e$ measured in optical wavelengths would become smaller because the light from star-forming galaxies in dominated by the disk component, which is more extended than the underlying mass profile. \citep{szo13,lan14,mos17}. Empirically, this change in size is found to be in general small \citep{szo13,mos17}. Overall, the newly-formed quiescent galaxies have masses and sizes similar to their progenitor star-forming galaxies and larger than pre-existing, older quiescent galaxies. 

On the other hand, mechanisms that shut off the star formation rapidly ($<$ a few hundred Myr) can produce quiescent galaxies much smaller than their progenitor star-forming galaxies. The mass increases due to intense star formation. The size, quantified as $R_e$, can become a few times smaller because new stars form in the center of the galaxy \citep{hop13,wel15,tac16}. The amount of change in mass and size depends on the properties of the progenitor \citep{hop13,dek14,blu18}.

\subsubsection{The relative importance of different routes}

The relative importance of different mechanisms likely depends on several factors. In the early universe, the merger rate is higher \citep{hop10,lot11,xu12,rod15,man16} and galaxies are gravitational unstable due to high gas fraction, rapid gas inflows and catastrophic events are more common \citep{dek09,dek14,zol15}. Empirically, the abundance of massive PSB galaxies increases along with redshifts up to $z\sim2$ \citep{tra04,yan09,ver10,wu14,wil16,mal18}, which also suggests that the star formation in galaxies is more often shut off rapidly at higher redshifts.

The environment is another relevant factor. The fraction of PSB galaxies is larger in galaxy groups and clusters \citep{tra03,pog09,muz12,dre13,wu14,pac18}. The merger rates is higher in denser regions \citep{sob11}. Moreover, in galaxy clusters, interstellar gas may be removed by interaction with dense intracluster medium, known as ram pressure stripping \citep{gun72}. The ram pressure is weak outside galaxy groups and clusters but is considered as an indispensable mechanism in order to explain the high PSB fractions in galaxy clusters \citep{muz14,wu14,pac18}. Only one PSB galaxy in our sample is a candidate cluster member \citep[according to the catalog of X-ray clusters in the COSMOS field,][]{fin07}. Our conclusion in this paper should apply to only field galaxies. 

In addition, at low stellar masses, galaxies are prone to environmental effect and AGN and stellar feedback. Extra quenching mechanisms may be in effect and produce low-mass PSB galaxies \citep{mal18}.

In reality, the evolutionary tracks of galaxies will not cleanly separate into slow and fast tracks but can fall anywhere in between. Regardless of the specific characteristics of any slow and fast tracks, our data unequivocally show that the relationship between star-formation history and structural evolution is complex and can vary strongly from one galaxy to another.

\subsection{Systematic uncertainties}

\subsubsection{The effect of metallicity}
\label{sec:met}
\begin{figure*}
	\centering
	\includegraphics[width=0.7\textwidth]{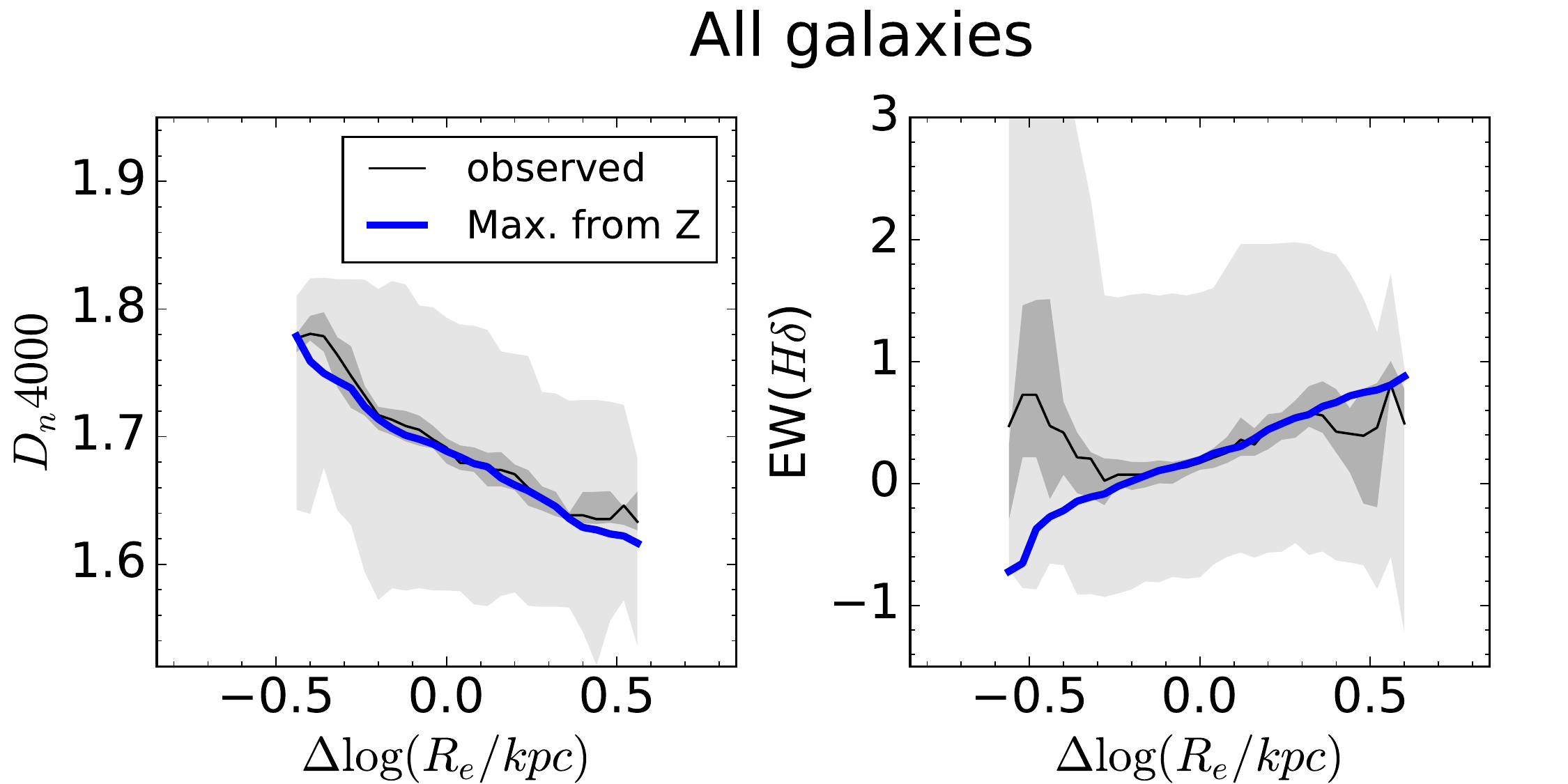}
	\caption{Testing the effect of metallicity on D$_n$4000 and EW(H$\delta$). The thin black lines and gray shaded areas are the same as Fig.~\ref{fig:AR}: the measured median indices, the uncertainties of the medians, and the 16th and 84th percentiles of the distributions. The thick blue lines are the D$_n$4000 and EW(H$\delta$) measured from mock spectra with a fixed age of 3~Gyr and a size-dependent metallicity $[Z/H] = -0.8 \times \Delta \log (R_e)$. The thick blue lines represent the maximum effect due to a size-dependent metallicity (see detailed discussion in Section~\ref{sec:met}). The metallicity produces comparable size dependence only in the most extreme case shown here. }
	\label{fig:met}
\end{figure*}

Both D$_n$4000 and EW(H$\delta$) depend on not only ages but also metallicities. In this paper, we interpret the dependence of indices on size as an effect of the age, assuming galaxies of the same masses having the same metallicity. In reality, the stellar mass-stellar metallicity relation of quiescent galaxies at $z\sim0.7$ shows a scatter of 0.16~dex \citep[][see also \citet{jor17}]{gal14}. The dependence of indices on size may alternatively be an effect of metallicity if the sizes and the metallicity are also correlated. Such a correlation has been observed in the local Universe \citep{mcd15,wu15,sco17,bar18,li18} but not yet at higher redshifts. 

Here we examine whether the variations in D$_n$4000 and EW(H$\delta$) can be attributed to variations in metallicity (instead of age). We test the most extreme scenario in which the scatter in the mass-size relation corresponds one-to-one to the scatter in the mass-stellar metallicity relation. 

We assign a fiducial metallicity to each galaxy based on the size:
\begin{equation}
[Z/H] = -0.8 \times \Delta \log(R_e)
\end{equation}
This is based on the assumption that the 0.16~dex scatter in the metallicity is entirely due to the variation in the size (0.2~dex scatter, Section~\ref{sec:ms}) and a galaxy with the median size has solar metallicity. We assign the same stellar age of 3~Gyr to all galaxies and produce mock spectra using \citet{bc03} SSP models then measure their D$_n$4000 and EW(H$\delta$). 

Fig.~\ref{fig:met} shows how D$_n$4000 and EW(H$\delta$) of the mock spectra vary with the size compared to the observed correlations. The similarity is striking, which immediately reveals that metallicity variation can play an important role. However, it is unlikely that the correlation between the size and the metallicity has no intrinsic scatter at all. The general trend we find in Section~\ref{sec:AR} that larger galaxies are younger than smaller galaxies is robust in a qualitative sense. But the metallicity dependence shows that the true correlation may be even weaker than what we find while assuming a fixed, solar metallicity. 

Assuming the same 3~Gyr population, the metallicity needs to be $[Z/H]<-1$ to have $\mbox{EW}(\mbox{H}\delta)>4$\AA, according to the model of \citet{tho11}. This metallicity is extremely low in comparison to the average ($[Z/H] \simeq 0.0\pm 0.16$) and has not yet been measured in galaxies at similar \citep{gal14,jor17} or even higher redshifts \citep[$z\sim3$,][]{som12,mar18}. It is thus unlikely that the strong Balmer absorption in PSB galaxies is entirely due to low metallicities.

\subsubsection{The effect of age gradient and slit loss}

Our spectra are obtained with slits of 1" width, which corresponds to $\sim7$~kpc at $z\sim0.7$. For large face-on galaxies or those misaligned with slits, the spectra miss light from galaxy outskirts, likely consists of younger stellar populations \citep{szo11,gon15,god17,wan17}. The age obtained from the spectra may be thus biased old.

We test the potential bias using a subset of spectra that is spatially-resolved. We select 152 larger quiescent galaxies ($R_e > 0.5"$) whose major axes align with the slits ($\Delta P.A. < 45\deg$) or face-on ($b/a>0.7$). We measure the D$_n$4000 and EW(H$\delta$) from the central 1" along the slit direction. The central D$_n$4000 is on average $0.05\pm0.01$ larger than integrating over the entire galaxy, suggesting the centers of galaxies are older. On the other hand, the difference in EW(H$\delta$) is consistent with zero, $0.02\pm0.03$\AA. We conclude that our result in Sec 3.2, namely that on average larger galaxies are younger is qualitatively not affected and in fact would be strengthened by the bias introduced by the finite slit width. 

We also confirm that the selection of PSB galaxies is not biased in the size; There is only one galaxy that would be defined as a PSB galaxy according to the integrated spectrum but not the central spectrum. Missing light at the outskirts of large galaxies does not result in a bias in the sizes of the PSB sample.

\subsubsection{Stellar mass-to-light ratio of PSB galaxies}

We measure the sizes using \textit{HST} F814W images, which corresponds to the rest-frame B-band for galaxies at $z\sim0.7$. The light of PSB galaxies is dominated by A-type stars, which may not necessarily trace the mass profiles well if the last episode of star-formation has distinct spatial distribution from the old stellar populations. 

A few works have studied the sizes of PSB galaxies using \textit{HST} F160W images but focusing on galaxies mainly at $z>1$, sampling the redder part at the rest-frame optical regime \citep{bel15,yan16,alm17}. These studies generally find that PSB galaxies are $\sim0.2$~dex smaller than average quiescent galaxies. A more reliable size should be measured at longer rest-frame wavelengths, preferentially in rest-frame near-IR. 
The new `drift and shift' observing mode of the \textit{HST} WFC3 IR channel allows a much faster survey speed \citep{mom17}. We have obtained the F160W images of a large fraction of our sample and will examine the structures of galaxies in multiple wavelengths. For PSB galaxies at $z>1$, it will require \textit{JWST} to provide images at longer wavelengths.

\section{Summary}
\label{sec:con}

In this paper, we present the size and age-sensitive spectral features (D$_n$4000 and EW(H$\delta$)) of 467 quiescent galaxies with stellar masses $M_\ast > 10^{10} M_\odot$ at $z\sim0.7$. We measure the sizes from \textit{HST} F814W images and age indicators using the ultra-deep spectra from the LEGA-C survey. The high quality spectra allows us to measure age indicators accurately for individual galaxies. 

At fixed stellar mass, the D$_n$4000 and EW(H$\delta$) show large individual variations, distributed over a wide range of $\mbox{D}_n4000 \simeq 1.6 - 1.8$ and $\mbox{EW}(\mbox{H}\delta) \simeq -1\AA - 2\AA$, respectively. These ranges correspond to an age range of $\sim1.5$~Gyr for a solar metallicity SSP. Nevertheless, we find a weak trend that larger galaxies have on average smaller D$_n$4000 and larger EW(H$\delta$). Excluding the most compact galaxies, the variation in median D$_n$4000 and EW(H$\delta$) is $\sim0.06$ and $\sim0.5$\AA, respectively. The corresponding age variation is $<500$~Myr. On the contrary, quiescent galaxies with rapid declining star-formation rates (PSB galaxies) do not follow the general age-size correlation; They are among the youngest and much smaller than other young quiescent galaxies.

These two seemingly contradictory statements are often presented separately in the literature and raise some dispute in the formation and evolution of quiescent galaxies. We demonstrate that both observations are correct and only with a large sample of galaxies with high-quality spectra can reveal the complex correlation. While the bulk population have the D$_n$4000 and EW(H$\delta$) suggesting that younger galaxies are on average larger, there is a small fraction of galaxies with elevated H$\delta$ absorption, indicating a rapid quenching star-formation history, that are small. 

This correlation suggests that there are multiple evolutionary pathways to quiescence. Star-forming galaxies that slowly exhaust their cold gas reservoirs join the large side of the red sequence. These galaxies do not go through a PSB phase because the star-formation fades in a longer timescale. They can drive at least partially the size evolution of quiescent galaxies and produce the progenitor bias. On the other hand, violent events such as galaxy mergers or violent disk instabilities can produce PSB galaxies that are much smaller than their star-forming progenitors and even smaller than the existing quiescent population. The star-formation stops within a few hundred Myrs, accompanied with structural changes that shrink the sizes of galaxies.

\acknowledgments
Based on observations made with ESO Telescopes at the La Silla Paranal Observatory under programme ID 194-A.2005 (The LEGA-C Public Spectroscopy Survey). We thank the referee for the valuable comments. This project has received funding from the European Research Council (ERC) under the European Union’s Horizon 2020 research and innovation programme (grant agreement No. 683184). CS acknowledges support from the Deutsche Forschungsemeinschaft (GZ: WE 4755/4-1).
VW acknowledges funding from the ERC (starting grant SEDmorph, PI. Wild)

\bibliography{age_size_v6.0.bbl}

\begin{thebibliography}{}
\expandafter\ifx\csname natexlab\endcsname\relax\def\natexlab#1{#1}\fi

\bibitem[{{Almaini} {et~al.}(2017){Almaini}, {Wild}, {Maltby}, {Hartley},
  {Simpson}, {Hatch}, {McLure}, {Dunlop}, \& {Rowlands}}]{alm17}
{Almaini}, O., {Wild}, V., {Maltby}, D.~T., {et~al.} 2017, \mnras, 472, 1401

\bibitem[{{Baldry} {et~al.}(2004){Baldry}, {Glazebrook}, {Brinkmann},
  {Ivezi{\'c}}, {Lupton}, {Nichol}, \& {Szalay}}]{bal04}
{Baldry}, I.~K., {Glazebrook}, K., {Brinkmann}, J., {et~al.} 2004, \apj, 600,
  681

\bibitem[{{Balogh} {et~al.}(1999){Balogh}, {Morris}, {Yee}, {Carlberg}, \&
  {Ellingson}}]{bal99}
{Balogh}, M.~L., {Morris}, S.~L., {Yee}, H.~K.~C., {Carlberg}, R.~G., \&
  {Ellingson}, E. 1999, \apj, 527, 54

\bibitem[{{Bari{\v s}i{\'c}} {et~al.}(2017){Bari{\v s}i{\'c}}, {van der Wel},
  {Bezanson}, {Pacifici}, {Noeske}, {Mu{\~n}oz-Mateos}, {Franx}, {Smol{\v
  c}i{\'c}}, {Bell}, {Brammer}, {Calhau}, {Chauk{\'e}}, {van Dokkum}, {van
  Houdt}, {Gallazzi}, {Labb{\'e}}, {Maseda}, {Muzzin}, {Sobral}, {Straatman},
  \& {Wu}}]{bar17}
{Bari{\v s}i{\'c}}, I., {van der Wel}, A., {Bezanson}, R., {et~al.} 2017, \apj,
  847, 72

\bibitem[{{Barnes} \& {Hernquist}(1996)}]{bar96}
{Barnes}, J.~E., \& {Hernquist}, L. 1996, \apj, 471, 115

\bibitem[{{Barnes} \& {Hernquist}(1991)}]{bar91}
{Barnes}, J.~E., \& {Hernquist}, L.~E. 1991, \apj, 370, L65

\bibitem[{{Barone} {et~al.}(2018){Barone}, {D'Eugenio}, {Colless}, {Scott},
  {van de Sande}, {Bland-Hawthorn}, {Brough}, {Bryant}, {Cortese}, {Croom},
  {Foster}, {Goodwin}, {Konstantopoulos}, {Lawrence}, {Lorente}, {Medling},
  {Owers}, \& {Richards}}]{bar18}
{Barone}, T.~M., {D'Eugenio}, F., {Colless}, M., {et~al.} 2018, ArXiv e-prints,
  arXiv:1802.04807

\bibitem[{{Barro} {et~al.}(2016){Barro}, {Kriek}, {P{\'e}rez-Gonz{\'a}lez},
  {Trump}, {Koo}, {Faber}, {Dekel}, {Primack}, {Guo}, {Kocevski},
  {Mu{\~n}oz-Mateos}, {Rujopakarn}, \& {Seth}}]{bar16}
{Barro}, G., {Kriek}, M., {P{\'e}rez-Gonz{\'a}lez}, P.~G., {et~al.} 2016,
  \apjl, 827, L32

\bibitem[{{Barro} {et~al.}(2017){Barro}, {Kriek}, {P{\'e}rez-Gonz{\'a}lez},
  {Diaz-Santos}, {Price}, {Rujopakarn}, {Pandya}, {Koo}, {Faber}, {Dekel},
  {Primack}, \& {Kocevski}}]{barr17}
---. 2017, \apjl, 851, L40

\bibitem[{{Bell} {et~al.}(2004){Bell}, {Wolf}, {Meisenheimer}, {Rix}, {Borch},
  {Dye}, {Kleinheinrich}, {Wisotzki}, \& {McIntosh}}]{bel04}
{Bell}, E.~F., {Wolf}, C., {Meisenheimer}, K., {et~al.} 2004, \apj, 608, 752

\bibitem[{{Belli} {et~al.}(2015){Belli}, {Newman}, \& {Ellis}}]{bel15}
{Belli}, S., {Newman}, A.~B., \& {Ellis}, R.~S. 2015, \apj, 799, 206

\bibitem[{{Best} {et~al.}(2005){Best}, {Kauffmann}, {Heckman}, {Brinchmann},
  {Charlot}, {Ivezi{\'c}}, \& {White}}]{bes05}
{Best}, P.~N., {Kauffmann}, G., {Heckman}, T.~M., {et~al.} 2005, \mnras, 362,
  25

\bibitem[{{Best} {et~al.}(2014){Best}, {Ker}, {Simpson}, {Rigby}, \&
  {Sabater}}]{bes14}
{Best}, P.~N., {Ker}, L.~M., {Simpson}, C., {Rigby}, E.~E., \& {Sabater}, J.
  2014, \mnras, 445, 955

\bibitem[{{Bezanson} {et~al.}(2018){Bezanson}, {van der Wel}, {Pacifici},
  {Noeske}, {Bari{\v{s}}i{\'c}}, {Bell}, {Brammer}, {Calhau}, {Chauke}, {van
  Dokkum}, {Franx}, {Gallazzi}, {van Houdt}, {Labb{\'e}}, {Maseda},
  {Mu{\~n}os-Mateos}, {Muzzin}, {van de Sande}, {Sobral}, {Straatman}, \&
  {Wu}}]{bez18}
{Bezanson}, R., {van der Wel}, A., {Pacifici}, C., {et~al.} 2018, ArXiv
  e-prints, arXiv:1804.02402

\bibitem[{{Bigiel} {et~al.}(2011){Bigiel}, {Leroy}, {Walter}, {Brinks}, {de
  Blok}, {Kramer}, {Rix}, {Schruba}, {Schuster}, {Usero}, \&
  {Wiesemeyer}}]{bie11}
{Bigiel}, F., {Leroy}, A.~K., {Walter}, F., {et~al.} 2011, \apjl, 730, L13

\bibitem[{{Blumenthal} \& {Barnes}(2018)}]{blu18}
{Blumenthal}, K.~A., \& {Barnes}, J.~E. 2018, \mnras, 1533

\bibitem[{{Bournaud} {et~al.}(2011){Bournaud}, {Chapon}, {Teyssier}, {Powell},
  {Elmegreen}, {Elmegreen}, {Duc}, {Contini}, {Epinat}, \& {Shapiro}}]{bou11}
{Bournaud}, F., {Chapon}, D., {Teyssier}, R., {et~al.} 2011, \apj, 730, 4

\bibitem[{{Bruzual} \& {Charlot}(2003)}]{bc03}
{Bruzual}, G., \& {Charlot}, S. 2003, \mnras, 344, 1000

\bibitem[{{Calzetti} {et~al.}(2000){Calzetti}, {Armus}, {Bohlin}, {Kinney},
  {Koornneef}, \& {Storchi-Bergmann}}]{cal00}
{Calzetti}, D., {Armus}, L., {Bohlin}, R.~C., {et~al.} 2000, \apj, 533, 682

\bibitem[{{Cappellari}(2017)}]{cap17}
{Cappellari}, M. 2017, \mnras, 466, 798

\bibitem[{{Cappellari} \& {Emsellem}(2004)}]{cap04}
{Cappellari}, M., \& {Emsellem}, E. 2004, \pasp, 116, 138

\bibitem[{{Cappellari} {et~al.}(2013){Cappellari}, {McDermid}, {Alatalo},
  {Blitz}, {Bois}, {Bournaud}, {Bureau}, {Crocker}, {Davies}, {Davis}, {de
  Zeeuw}, {Duc}, {Emsellem}, {Khochfar}, {Krajnovi{\'c}}, {Kuntschner},
  {Morganti}, {Naab}, {Oosterloo}, {Sarzi}, {Scott}, {Serra}, {Weijmans}, \&
  {Young}}]{cap13}
{Cappellari}, M., {McDermid}, R.~M., {Alatalo}, K., {et~al.} 2013, \mnras, 432,
  1862

\bibitem[{{Carollo} {et~al.}(2013){Carollo}, {Bschorr}, {Renzini}, {Lilly},
  {Capak}, {Cibinel}, {Ilbert}, {Onodera}, {Scoville}, {Cameron}, {Mobasher},
  {Sanders}, \& {Taniguchi}}]{car13}
{Carollo}, C.~M., {Bschorr}, T.~J., {Renzini}, A., {et~al.} 2013, \apj, 773,
  112

\bibitem[{{Chabrier}(2003)}]{cha03}
{Chabrier}, G. 2003, \pasp, 115, 763

\bibitem[{{Chauke} {et~al.}(2018){Chauke}, {van der Wel}, {Pacifici},
  {Bezanson}, {Wu}, {Gallazzi}, {Noeske}, {Straatman}, {Mu{\~n}os-Mateos},
  {Franx}, {Bari{\v{s}}i{\'c}}, {Bell}, {Brammer}, {Calhau}, {van Houdt},
  {Labb{\'e}}, {Maseda}, {Muzzin}, {Rix}, \& {Sobral}}]{cha18}
{Chauke}, P., {van der Wel}, A., {Pacifici}, C., {et~al.} 2018, ArXiv e-prints,
  arXiv:1805.02568

\bibitem[{Cleveland \& Devlin(1988)}]{cle88}
Cleveland, W.~S., \& Devlin, S.~J. 1988, Journal of the American Statistical
  Association, 83, 596

\bibitem[{{Croton} {et~al.}(2006){Croton}, {Springel}, {White}, {De Lucia},
  {Frenk}, {Gao}, {Jenkins}, {Kauffmann}, {Navarro}, \& {Yoshida}}]{cro06}
{Croton}, D.~J., {Springel}, V., {White}, S. D.~M., {et~al.} 2006, \mnras, 365,
  11

\bibitem[{{Dalcanton} {et~al.}(1997){Dalcanton}, {Spergel}, \&
  {Summers}}]{dal97}
{Dalcanton}, J.~J., {Spergel}, D.~N., \& {Summers}, F.~J. 1997, \apj, 482, 659

\bibitem[{{Dekel} \& {Birnboim}(2006)}]{dek06}
{Dekel}, A., \& {Birnboim}, Y. 2006, \mnras, 368, 2

\bibitem[{{Dekel} \& {Burkert}(2014)}]{dek14}
{Dekel}, A., \& {Burkert}, A. 2014, \mnras, 438, 1870

\bibitem[{{Dekel} {et~al.}(2009){Dekel}, {Birnboim}, {Engel}, {Freundlich},
  {Goerdt}, {Mumcuoglu}, {Neistein}, {Pichon}, {Teyssier}, \& {Zinger}}]{dek09}
{Dekel}, A., {Birnboim}, Y., {Engel}, G., {et~al.} 2009, \nat, 457, 451

\bibitem[{{Dressler} \& {Gunn}(1983)}]{dre83}
{Dressler}, A., \& {Gunn}, J.~E. 1983, \apj, 270, 7

\bibitem[{{Dressler} {et~al.}(2013){Dressler}, {Oemler}, {Poggianti},
  {Gladders}, {Abramson}, \& {Vulcani}}]{dre13}
{Dressler}, A., {Oemler}, Augustus, J., {Poggianti}, B.~M., {et~al.} 2013,
  \apj, 770, 62

\bibitem[{{Dressler} {et~al.}(1999){Dressler}, {Smail}, {Poggianti}, {Butcher},
  {Couch}, {Ellis}, \& {Oemler}}]{dre99}
{Dressler}, A., {Smail}, I., {Poggianti}, B.~M., {et~al.} 1999, The
  Astrophysical Journal Supplement Series, 122, 51

\bibitem[{{Efstathiou}(2000)}]{efs00}
{Efstathiou}, G. 2000, \mnras, 317, 697

\bibitem[{{Ellison} {et~al.}(2018){Ellison}, {Catinella}, \& {Cortese}}]{ell18}
{Ellison}, S.~L., {Catinella}, B., \& {Cortese}, L. 2018, \mnras, 1184

\bibitem[{{Fabian}(2012)}]{fab12}
{Fabian}, A.~C. 2012, Annual Review of Astronomy and Astrophysics, 50, 455

\bibitem[{{Fagioli} {et~al.}(2016){Fagioli}, {Carollo}, {Renzini}, {Lilly},
  {Onodera}, \& {Tacchella}}]{fag16}
{Fagioli}, M., {Carollo}, C.~M., {Renzini}, A., {et~al.} 2016, \apj, 831, 173

\bibitem[{{Finoguenov} {et~al.}(2007){Finoguenov}, {Guzzo}, {Hasinger},
  {Scoville}, {Aussel}, {B{\"o}hringer}, {Brusa}, {Capak}, {Cappelluti},
  {Comastri}, {Giodini}, {Griffiths}, {Impey}, {Koekemoer}, {Kneib},
  {Leauthaud}, {Le F{\`e}vre}, {Lilly}, {Mainieri}, {Massey}, {McCracken},
  {Mobasher}, {Murayama}, {Peacock}, {Sakelliou}, {Schinnerer}, {Silverman},
  {Smol{\v{c}}i{\'c}}, {Taniguchi}, {Tasca}, {Taylor}, {Trump}, \&
  {Zamorani}}]{fin07}
{Finoguenov}, A., {Guzzo}, L., {Hasinger}, G., {et~al.} 2007, The Astrophysical
  Journal Supplement Series, 172, 182

\bibitem[{{Franzetti} {et~al.}(2007){Franzetti}, {Scodeggio}, {Garilli},
  {Vergani}, {Maccagni}, {Guzzo}, {Tresse}, {Ilbert}, {Lamareille}, {Contini},
  {Le F{\`e}vre}, {Zamorani}, {Brinchmann}, {Charlot}, {Bottini}, {Le Brun},
  {Picat}, {Scaramella}, {Vettolani}, {Zanichelli}, {Adami}, {Arnouts},
  {Bardelli}, {Bolzonella}, {Cappi}, {Ciliegi}, {Foucaud}, {Gavignaud},
  {Iovino}, {McCracken}, {Marano}, {Marinoni}, {Mazure}, {Meneux}, {Merighi},
  {Paltani}, {Pell{\`o}}, {Pollo}, {Pozzetti}, {Radovich}, {Zucca}, {Cucciati},
  \& {Walcher}}]{fra07}
{Franzetti}, P., {Scodeggio}, M., {Garilli}, B., {et~al.} 2007, \aap, 465, 711

\bibitem[{{Gabor} \& {Bournaud}(2013)}]{gab11}
{Gabor}, J.~M., \& {Bournaud}, F. 2013, \mnras, 434, 606

\bibitem[{{Gallazzi} {et~al.}(2014){Gallazzi}, {Bell}, {Zibetti}, {Brinchmann},
  \& {Kelson}}]{gal14}
{Gallazzi}, A., {Bell}, E.~F., {Zibetti}, S., {Brinchmann}, J., \& {Kelson},
  D.~D. 2014, \apj, 788, 72

\bibitem[{{Genel} {et~al.}(2018){Genel}, {Nelson}, {Pillepich}, {Springel},
  {Pakmor}, {Weinberger}, {Hernquist}, {Naiman}, {Vogelsberger}, {Marinacci},
  \& {Torrey}}]{gen18}
{Genel}, S., {Nelson}, D., {Pillepich}, A., {et~al.} 2018, \mnras, 474, 3976

\bibitem[{{Genzel} {et~al.}(2015){Genzel}, {Tacconi}, {Lutz}, {Saintonge},
  {Berta}, {Magnelli}, {Combes}, {Garc{\'{\i}}a-Burillo}, {Neri}, {Bolatto},
  {Contini}, {Lilly}, {Boissier}, {Boone}, {Bouch{\'e}}, {Bournaud}, {Burkert},
  {Carollo}, {Colina}, {Cooper}, {Cox}, {Feruglio}, {F{\"o}rster Schreiber},
  {Freundlich}, {Gracia-Carpio}, {Juneau}, {Kovac}, {Lippa}, {Naab}, {Salome},
  {Renzini}, {Sternberg}, {Walter}, {Weiner}, {Weiss}, \& {Wuyts}}]{gen15}
{Genzel}, R., {Tacconi}, L.~J., {Lutz}, D., {et~al.} 2015, \apj, 800, 20

\bibitem[{{Goddard} {et~al.}(2017){Goddard}, {Thomas}, {Maraston}, {Westfall},
  {Etherington}, {Riffel}, {Mallmann}, {Zheng}, {Argudo-Fern{\'a}ndez}, {Lian},
  {Bershady}, {Bundy}, {Drory}, {Law}, {Yan}, {Wake}, {Weijmans}, {Bizyaev},
  {Brownstein}, {Lane}, {Maiolino}, {Masters}, {Merrifield}, {Nitschelm},
  {Pan}, {Roman-Lopes}, {Storchi-Bergmann}, \& {Schneider}}]{god17}
{Goddard}, D., {Thomas}, D., {Maraston}, C., {et~al.} 2017, \mnras, 466, 4731

\bibitem[{{Gonz{\'a}lez Delgado} {et~al.}(2015){Gonz{\'a}lez Delgado},
  {Garc{\'{\i}}a-Benito}, {P{\'e}rez}, {Cid Fernandes}, {de Amorim},
  {Cortijo-Ferrero}, {Lacerda}, {L{\'o}pez Fern{\'a}ndez}, {Vale-Asari},
  {S{\'a}nchez}, {Moll{\'a}}, {Ruiz-Lara}, {S{\'a}nchez-Bl{\'a}zquez},
  {Walcher}, {Alves}, {Aguerri}, {Bekerait{\'e}}, {Bland-Hawthorn}, {Galbany},
  {Gallazzi}, {Husemann}, {Iglesias-P{\'a}ramo}, {Kalinova},
  {L{\'o}pez-S{\'a}nchez}, {Marino}, {M{\'a}rquez}, {Masegosa}, {Mast},
  {M{\'e}ndez-Abreu}, {Mendoza}, {del Olmo}, {P{\'e}rez}, {Quirrenbach}, \&
  {Zibetti}}]{gon15}
{Gonz{\'a}lez Delgado}, R.~M., {Garc{\'{\i}}a-Benito}, R., {P{\'e}rez}, E.,
  {et~al.} 2015, \aap, 581, A103

\bibitem[{{Gunn} \& {Gott}(1972)}]{gun72}
{Gunn}, J.~E., \& {Gott}, J.~Richard, I. 1972, \apj, 176, 1

\bibitem[{{Haines} {et~al.}(2017){Haines}, {Iovino}, {Krywult}, {Guzzo},
  {Davidzon}, {Bolzonella}, {Garilli}, {Scodeggio}, {Granett}, {de la Torre},
  {De Lucia}, {Abbas}, {Adami}, {Arnouts}, {Bottini}, {Cappi}, {Cucciati},
  {Franzetti}, {Fritz}, {Gargiulo}, {Le Brun}, {Le F{\`e}vre}, {Maccagni},
  {Ma{\l}ek}, {Marulli}, {Moutard}, {Polletta}, {Pollo}, {Tasca}, {Tojeiro},
  {Vergani}, {Zanichelli}, {Zamorani}, {Bel}, {Branchini}, {Coupon}, {Ilbert},
  {Moscardini}, {Peacock}, \& {Siudek}}]{hai17}
{Haines}, C.~P., {Iovino}, A., {Krywult}, J., {et~al.} 2017, \aap, 605, A4

\bibitem[{{Hopkins} {et~al.}(2013){Hopkins}, {Cox}, {Hernquist}, {Narayanan},
  {Hayward}, \& {Murray}}]{hop13}
{Hopkins}, P.~F., {Cox}, T.~J., {Hernquist}, L., {et~al.} 2013, \mnras, 430,
  1901

\bibitem[{{Hopkins} {et~al.}(2006){Hopkins}, {Hernquist}, {Cox}, {Di Matteo},
  {Robertson}, \& {Springel}}]{hop06}
{Hopkins}, P.~F., {Hernquist}, L., {Cox}, T.~J., {et~al.} 2006, The
  Astrophysical Journal Supplement Series, 163, 1

\bibitem[{{Hopkins} {et~al.}(2008){Hopkins}, {Hernquist}, {Cox}, \& {Kere{\v
  s}}}]{hop08}
{Hopkins}, P.~F., {Hernquist}, L., {Cox}, T.~J., \& {Kere{\v s}}, D. 2008,
  \apjs, 175, 356

\bibitem[{{Hopkins} {et~al.}(2010){Hopkins}, {Bundy}, {Croton}, {Hernquist},
  {Keres}, {Khochfar}, {Stewart}, {Wetzel}, \& {Younger}}]{hop10}
{Hopkins}, P.~F., {Bundy}, K., {Croton}, D., {et~al.} 2010, \apj, 715, 202

\bibitem[{{Huang} {et~al.}(2012){Huang}, {Haynes}, {Giovanelli}, \&
  {Brinchmann}}]{hua12}
{Huang}, S., {Haynes}, M.~P., {Giovanelli}, R., \& {Brinchmann}, J. 2012, \apj,
  756, 113

\bibitem[{{J{\o}rgensen} {et~al.}(2017){J{\o}rgensen}, {Chiboucas}, {Berkson},
  {Smith}, {Takamiya}, \& {Villaume}}]{jor17}
{J{\o}rgensen}, I., {Chiboucas}, K., {Berkson}, E., {et~al.} 2017, \aj, 154,
  251

\bibitem[{{Kauffmann} {et~al.}(2003){Kauffmann}, {Heckman}, {White}, {Charlot},
  {Tremonti}, {Brinchmann}, {Bruzual}, {Peng}, {Seibert}, {Bernardi},
  {Blanton}, {Brinkmann}, {Castander}, {Cs{\'a}bai}, {Fukugita}, {Ivezic},
  {Munn}, {Nichol}, {Padmanabhan}, {Thakar}, {Weinberg}, \& {York}}]{kau03a}
{Kauffmann}, G., {Heckman}, T.~M., {White}, S.~D.~M., {et~al.} 2003, \mnras,
  341, 33

\bibitem[{{Kaviraj} {et~al.}(2007){Kaviraj}, {Kirkby}, {Silk}, \&
  {Sarzi}}]{kav07}
{Kaviraj}, S., {Kirkby}, L.~A., {Silk}, J., \& {Sarzi}, M. 2007, \mnras, 382,
  960

\bibitem[{{Keating} {et~al.}(2015){Keating}, {Abraham}, {Schiavon}, {Graves},
  {Damjanov}, {Yan}, {Newman}, \& {Simard}}]{kea15}
{Keating}, S.~K., {Abraham}, R.~G., {Schiavon}, R., {et~al.} 2015, \apj, 798,
  26

\bibitem[{{Kere{\v{s}}} {et~al.}(2005){Kere{\v{s}}}, {Katz}, {Weinberg}, \&
  {Dav{\'e}}}]{ker05}
{Kere{\v{s}}}, D., {Katz}, N., {Weinberg}, D.~H., \& {Dav{\'e}}, R. 2005,
  \mnras, 363, 2

\bibitem[{{Kriek} {et~al.}(2009){Kriek}, {van Dokkum}, {Labb{\'e}}, {Franx},
  {Illingworth}, {Marchesini}, \& {Quadri}}]{kri09}
{Kriek}, M., {van Dokkum}, P.~G., {Labb{\'e}}, I., {et~al.} 2009, \apj, 700,
  221

\bibitem[{{Kriek} {et~al.}(2010){Kriek}, {Labb{\'e}}, {Conroy}, {Whitaker},
  {van Dokkum}, {Brammer}, {Franx}, {Illingworth}, {Marchesini}, {Muzzin},
  {Quadri}, \& {Rudnick}}]{kri10}
{Kriek}, M., {Labb{\'e}}, I., {Conroy}, C., {et~al.} 2010, \apj, 722, L64

\bibitem[{{Lang} {et~al.}(2014){Lang}, {Wuyts}, {Somerville}, {F{\"o}rster
  Schreiber}, {Genzel}, {Bell}, {Brammer}, {Dekel}, {Faber}, {Ferguson},
  {Grogin}, {Kocevski}, {Koekemoer}, {Lutz}, {McGrath}, {Momcheva}, {Nelson},
  {Primack}, {Rosario}, {Skelton}, {Tacconi}, {van Dokkum}, \&
  {Whitaker}}]{lan14}
{Lang}, P., {Wuyts}, S., {Somerville}, R.~S., {et~al.} 2014, \apj, 788, 11

\bibitem[{{Le Borgne} {et~al.}(2006){Le Borgne}, {Abraham}, {Daniel},
  {McCarthy}, {Glazebrook}, {Savaglio}, {Crampton}, {Juneau}, {Carlberg},
  {Chen}, {Marzke}, {Roth}, {J{\o}rgensen}, \& {Murowinski}}]{leb06}
{Le Borgne}, D., {Abraham}, R., {Daniel}, K., {et~al.} 2006, \apj, 642, 48

\bibitem[{{Le F{\`e}vre} {et~al.}(2003){Le F{\`e}vre}, {Saisse}, {Mancini},
  {Brau-Nogue}, {Caputi}, {Castinel}, {D'Odorico}, {Garilli}, {Kissler-Patig},
  {Lucuix}, {Mancini}, {Pauget}, {Sciarretta}, {Scodeggio}, {Tresse}, \&
  {Vettolani}}]{lef03}
{Le F{\`e}vre}, O., {Saisse}, M., {Mancini}, D., {et~al.} 2003, in \procspie,
  Vol. 4841, Instrument Design and Performance for Optical/Infrared
  Ground-based Telescopes, ed. M.~{Iye} \& A.~F.~M. {Moorwood}, 1670--1681

\bibitem[{{Lemaux} {et~al.}(2010){Lemaux}, {Lubin}, {Shapley}, {Kocevski},
  {Gal}, \& {Squires}}]{lem10}
{Lemaux}, B.~C., {Lubin}, L.~M., {Shapley}, A., {et~al.} 2010, \apj, 716, 970

\bibitem[{{Lemaux} {et~al.}(2017){Lemaux}, {Tomczak}, {Lubin}, {Wu}, {Gal},
  {Rumbaugh}, {Kocevski}, \& {Squires}}]{lem17}
{Lemaux}, B.~C., {Tomczak}, A.~R., {Lubin}, L.~M., {et~al.} 2017, \mnras, 472,
  419

\bibitem[{{Leroy} {et~al.}(2008){Leroy}, {Walter}, {Brinks}, {Bigiel}, {de
  Blok}, {Madore}, \& {Thornley}}]{ler08}
{Leroy}, A.~K., {Walter}, F., {Brinks}, E., {et~al.} 2008, \aj, 136, 2782

\bibitem[{{Li} {et~al.}(2018){Li}, {Mao}, {Cappellari}, {Ge}, {Long}, {Li},
  {Mo}, {Li}, {Zheng}, {Bundy}, {Thomas}, {Brownstein}, {Lopes}, {Law}, \&
  {Drory}}]{li18}
{Li}, H., {Mao}, S., {Cappellari}, M., {et~al.} 2018, ArXiv e-prints,
  arXiv:1802.01819

\bibitem[{{Lilly} \& {Carollo}(2016)}]{lil16}
{Lilly}, S.~J., \& {Carollo}, C.~M. 2016, \apj, 833, 1

\bibitem[{{Lotz} {et~al.}(2011){Lotz}, {Jonsson}, {Cox}, {Croton}, {Primack},
  {Somerville}, \& {Stewart}}]{lot11}
{Lotz}, J.~M., {Jonsson}, P., {Cox}, T.~J., {et~al.} 2011, \apj, 742, 103

\bibitem[{{Maltby} {et~al.}(2018){Maltby}, {Almaini}, {Wild}, {Hatch},
  {Hartley}, {Simpson}, {Rowlands}, \& {Socolovsky}}]{mal18}
{Maltby}, D.~T., {Almaini}, O., {Wild}, V., {et~al.} 2018, \mnras, 480, 381

\bibitem[{{Man} {et~al.}(2016){Man}, {Zirm}, \& {Toft}}]{man16}
{Man}, A.~W.~S., {Zirm}, A.~W., \& {Toft}, S. 2016, \apj, 830, 89

\bibitem[{{Marques-Chaves} {et~al.}(2018){Marques-Chaves}, {P{\'e}rez-Fournon},
  {Gav azzi}, {Mart{\'\i}nez-Navajas}, {Riechers}, {Rigopoulou},
  {Cabrera-Lavers}, {Clements}, {Cooray}, {Farrah}, {Ivison},
  {Jim{\'e}nez-{\'A}ngel}, {Nayyeri}, {Oliver}, {Omont}, {Scott}, {Shu}, \&
  {Wardlow}}]{mar18}
{Marques-Chaves}, R., {P{\'e}rez-Fournon}, I., {Gav azzi}, R., {et~al.} 2018,
  \apj, 854, 151

\bibitem[{{McDermid} {et~al.}(2015){McDermid}, {Alatalo}, {Blitz}, {Bournaud},
  {Bureau}, {Cappellari}, {Crocker}, {Davies}, {Davis}, {de Zeeuw}, {Duc},
  {Emsellem}, {Khochfar}, {Krajnovi{\'c}}, {Kuntschner}, {Morganti}, {Naab},
  {Oosterloo}, {Sarzi}, {Scott}, {Serra}, {Weijmans}, \& {Young}}]{mcd15}
{McDermid}, R.~M., {Alatalo}, K., {Blitz}, L., {et~al.} 2015, \mnras, 448, 3484

\bibitem[{{Mihos} \& {Hernquist}(1994)}]{mih94}
{Mihos}, J.~C., \& {Hernquist}, L. 1994, \apjl, 437, L47

\bibitem[{{Mihos} {et~al.}(1997){Mihos}, {McGaugh}, \& {de Blok}}]{mih97}
{Mihos}, J.~C., {McGaugh}, S.~S., \& {de Blok}, W.~J.~G. 1997, \apjl, 477, L79

\bibitem[{{Momcheva} {et~al.}(2017){Momcheva}, {van Dokkum}, {van der Wel},
  {Brammer}, {MacKenty}, {Nelson}, {Leja}, {Muzzin}, \& {Franx}}]{mom17}
{Momcheva}, I.~G., {van Dokkum}, P.~G., {van der Wel}, A., {et~al.} 2017,
  Publications of the Astronomical Society of the Pacific, 129, 015004

\bibitem[{{Moresco} {et~al.}(2013){Moresco}, {Pozzetti}, {Cimatti}, {Zamorani},
  {Bolzonella}, {Lamareille}, {Mignoli}, {Zucca}, {Lilly}, {Carollo},
  {Contini}, {Kneib}, {Le F{\`e}vre}, {Mainieri}, {Renzini}, {Scodeggio},
  {Bardelli}, {Bongiorno}, {Caputi}, {Cucciati}, {de la Torre}, {de Ravel},
  {Franzetti}, {Garilli}, {Iovino}, {Kampczyk}, {Knobel}, {Kova{\v c}}, {Le
  Borgne}, {Le Brun}, {Maier}, {Pell{\'o}}, {Peng}, {Perez-Montero},
  {Presotto}, {Silverman}, {Tanaka}, {Tasca}, {Tresse}, {Vergani}, {Barnes},
  {Bordoloi}, {Cappi}, {Diener}, {Koekemoer}, {Le Floc'h}, {L{\'o}pez-Sanjuan},
  {McCracken}, {Nair}, {Oesch}, {Scarlata}, {Scoville}, \& {Welikala}}]{mor13}
{Moresco}, M., {Pozzetti}, L., {Cimatti}, A., {et~al.} 2013, \aap, 558, A61

\bibitem[{{Mosleh} {et~al.}(2017){Mosleh}, {Tacchella}, {Renzini}, {Carollo},
  {Molaeinezhad}, {Onodera}, {Khosroshahi}, \& {Lilly}}]{mos17}
{Mosleh}, M., {Tacchella}, S., {Renzini}, A., {et~al.} 2017, \apj, 837, 2

\bibitem[{{Muzzin} {et~al.}(2012){Muzzin}, {Wilson}, {Yee}, {Gilbank},
  {Hoekstra}, {Demarco}, {Balogh}, {van Dokkum}, {Franx}, {Ellingson}, {Hicks},
  {Nantais}, {Noble}, {Lacy}, {Lidman}, {Rettura}, {Surace}, \& {Webb}}]{muz12}
{Muzzin}, A., {Wilson}, G., {Yee}, H.~K.~C., {et~al.} 2012, \apj, 746, 188

\bibitem[{{Muzzin} {et~al.}(2013){Muzzin}, {Marchesini}, {Stefanon}, {Franx},
  {McCracken}, {Milvang-Jensen}, {Dunlop}, {Fynbo}, {Brammer}, {Labb{\'e}}, \&
  {van Dokkum}}]{muz13}
{Muzzin}, A., {Marchesini}, D., {Stefanon}, M., {et~al.} 2013, \apj, 777, 18

\bibitem[{{Muzzin} {et~al.}(2014){Muzzin}, {van der Burg}, {McGee}, {Balogh},
  {Franx}, {Hoekstra}, {Hudson}, {Noble}, {Taranu}, {Webb}, {Wilson}, \&
  {Yee}}]{muz14}
{Muzzin}, A., {van der Burg}, R.~F.~J., {McGee}, S.~L., {et~al.} 2014, \apj,
  796, 65

\bibitem[{{Paccagnella} {et~al.}(2018){Paccagnella}, {Vulcani}, {Poggianti},
  {Moretti}, {Fritz}, \& {Fasano}}]{pac18}
{Paccagnella}, A., {Vulcani}, B., {Poggianti}, B.~M., {et~al.} 2018, ArXiv
  e-prints, arXiv:1805.11475

\bibitem[{{Paulino-Afonso} {et~al.}(2017){Paulino-Afonso}, {Sobral},
  {Buitrago}, \& {Afonso}}]{pau17}
{Paulino-Afonso}, A., {Sobral}, D., {Buitrago}, F., \& {Afonso}, J. 2017,
  \mnras, 465, 2717

\bibitem[{{Peng} {et~al.}(2010){Peng}, {Ho}, {Impey}, \& {Rix}}]{pen10}
{Peng}, C.~Y., {Ho}, L.~C., {Impey}, C.~D., \& {Rix}, H.-W. 2010, \aj, 139,
  2097

\bibitem[{{Pillepich} {et~al.}(2018){Pillepich}, {Springel}, {Nelson}, {Genel},
  {Naiman}, {Pakmor}, {Hernquist}, {Torrey}, {Vogelsberger}, {Weinberger}, \&
  {Marinacci}}]{pil18}
{Pillepich}, A., {Springel}, V., {Nelson}, D., {et~al.} 2018, \mnras, 473, 4077

\bibitem[{{Poggianti} {et~al.}(2009){Poggianti}, {Arag{\'o}n-Salamanca},
  {Zaritsky}, {De Lucia}, {Milvang- Jensen}, {Desai}, {Jablonka}, {Halliday},
  {Rudnick}, {Varela}, {Bamford}, {Best}, {Clowe}, {Noll}, {Saglia},
  {Pell{\'o}}, {Simard}, {von der Linden}, \& {White}}]{pog09}
{Poggianti}, B.~M., {Arag{\'o}n-Salamanca}, A., {Zaritsky}, D., {et~al.} 2009,
  \apj, 693, 112

\bibitem[{{Poggianti} {et~al.}(2013){Poggianti}, {Calvi}, {Bindoni},
  {D'Onofrio}, {Moretti}, {Valentinuzzi}, {Fasano}, {Fritz}, {De Lucia},
  {Vulcani}, {Bettoni}, {Gullieuszik}, \& {Omizzolo}}]{pog13}
{Poggianti}, B.~M., {Calvi}, R., {Bindoni}, D., {et~al.} 2013, \apj, 762, 77

\bibitem[{{Popping} {et~al.}(2015){Popping}, {Caputi}, {Trager}, {Somerville},
  {Dekel}, {Kassin}, {Kocevski}, {Koekemoer}, {Faber}, {Ferguson}, {Galametz},
  {Grogin}, {Guo}, {Lu}, {Wel}, \& {Weiner}}]{pop15}
{Popping}, G., {Caputi}, K.~I., {Trager}, S.~C., {et~al.} 2015, \mnras, 454,
  2258

\bibitem[{{Popping} {et~al.}(2017){Popping}, {Decarli}, {Man}, {Nelson},
  {B{\'e}thermin}, {De Breuck}, {Mainieri}, {van Dokkum}, {Gullberg}, {van
  Kampen}, {Spaans}, \& {Trager}}]{pop17}
{Popping}, G., {Decarli}, R., {Man}, A. W.~S., {et~al.} 2017, \aap, 602, A11

\bibitem[{{Rodriguez-Gomez} {et~al.}(2015){Rodriguez-Gomez}, {Genel},
  {Vogelsberger}, {Sijacki}, {Pillepich}, {Sales}, {Torrey}, {Snyder},
  {Nelson}, {Springel}, {Ma}, \& {Hernquist}}]{rod15}
{Rodriguez-Gomez}, V., {Genel}, S., {Vogelsberger}, M., {et~al.} 2015, \mnras,
  449, 49

\bibitem[{{Saintonge} {et~al.}(2011){Saintonge}, {Kauffmann}, {Wang}, {Kramer},
  {Tacconi}, {Buchbender}, {Catinella}, {Graci{\'a}-Carpio}, {Cortese},
  {Fabello}, {Fu}, {Genzel}, {Giovanelli}, {Guo}, {Haynes}, {Heckman},
  {Krumholz}, {Lemonias}, {Li}, {Moran}, {Rodriguez-Fernandez}, {Schiminovich},
  {Schuster}, \& {Sievers}}]{sai11}
{Saintonge}, A., {Kauffmann}, G., {Wang}, J., {et~al.} 2011, \mnras, 415, 61

\bibitem[{{Schawinski} {et~al.}(2014){Schawinski}, {Urry}, {Simmons},
  {Fortson}, {Kaviraj}, {Keel}, {Lintott}, {Masters}, {Nichol}, {Sarzi},
  {Skibba}, {Treister}, {Willett}, {Wong}, \& {Yi}}]{sch14}
{Schawinski}, K., {Urry}, C.~M., {Simmons}, B.~D., {et~al.} 2014, \mnras, 440,
  889

\bibitem[{{Schiminovich} {et~al.}(2010){Schiminovich}, {Catinella},
  {Kauffmann}, {Fabello}, {Wang}, {Hummels}, {Lemonias}, {Moran}, {Wu},
  {Giovanelli}, {Haynes}, {Heckman}, {Basu-Zych}, {Blanton}, {Brinchmann},
  {Budav{\'a}ri}, {Gon{\c c}alves}, {Johnson}, {Kennicutt}, {Madore}, {Martin},
  {Rich}, {Tacconi}, {Thilker}, {Wild}, \& {Wyder}}]{sch10}
{Schiminovich}, D., {Catinella}, B., {Kauffmann}, G., {et~al.} 2010, \mnras,
  408, 919

\bibitem[{{Schruba} {et~al.}(2011){Schruba}, {Leroy}, {Walter}, {Bigiel},
  {Brinks}, {de Blok}, {Dumas}, {Kramer}, {Rosolowsky}, {Sandstrom},
  {Schuster}, {Usero}, {Weiss}, \& {Wiesemeyer}}]{sch11}
{Schruba}, A., {Leroy}, A.~K., {Walter}, F., {et~al.} 2011, \aj, 142, 37

\bibitem[{{Scott} {et~al.}(2017){Scott}, {Brough}, {Croom}, {Davies}, {van de
  Sande}, {Allen}, {Bland-Hawthorn}, {Bryant}, {Cortese}, {D'Eugenio},
  {Federrath}, {Ferreras}, {Goodwin}, {Groves}, {Konstantopoulos}, {Lawrence},
  {Medling}, {Moffett}, {Owers}, {Richards}, {Robotham}, {Tonini}, \&
  {Yi}}]{sco17}
{Scott}, N., {Brough}, S., {Croom}, S.~M., {et~al.} 2017, \mnras, 472, 2833

\bibitem[{{Scoville} {et~al.}(2007){Scoville}, {Abraham}, {Aussel}, {Barnes},
  {Benson}, {Blain}, {Calzetti}, {Comastri}, {Capak}, {Carilli}, {Carlstrom},
  {Carollo}, {Colbert}, {Daddi}, {Ellis}, {Elvis}, {Ewald}, {Fall},
  {Franceschini}, {Giavalisco}, {Green}, {Griffiths}, {Guzzo}, {Hasinger},
  {Impey}, {Kneib}, {Koda}, {Koekemoer}, {Lefevre}, {Lilly}, {Liu},
  {McCracken}, {Massey}, {Mellier}, {Miyazaki}, {Mobasher}, {Mould}, {Norman},
  {Refregier}, {Renzini}, {Rhodes}, {Rich}, {Sanders}, {Schiminovich},
  {Schinnerer}, {Scodeggio}, {Sheth}, {Shopbell}, {Taniguchi}, {Tyson}, {Urry},
  {Van Waerbeke}, {Vettolani}, {White}, \& {Yan}}]{sco07}
{Scoville}, N., {Abraham}, R.~G., {Aussel}, H., {et~al.} 2007, \apjs, 172, 38

\bibitem[{{Shankar} {et~al.}(2010){Shankar}, {Marulli}, {Bernardi}, {Dai},
  {Hyde}, \& {Sheth}}]{sha10a}
{Shankar}, F., {Marulli}, F., {Bernardi}, M., {et~al.} 2010, \mnras, 403, 117

\bibitem[{{Shen} {et~al.}(2003){Shen}, {Mo}, {White}, {Blanton}, {Kauffmann},
  {Voges}, {Brinkmann}, \& {Csabai}}]{she03}
{Shen}, S., {Mo}, H.~J., {White}, S.~D.~M., {et~al.} 2003, \mnras, 343, 978

\bibitem[{{Sijacki} {et~al.}(2007){Sijacki}, {Springel}, {Di Matteo}, \&
  {Hernquist}}]{sij07}
{Sijacki}, D., {Springel}, V., {Di Matteo}, T., \& {Hernquist}, L. 2007,
  \mnras, 380, 877

\bibitem[{{Siudek} {et~al.}(2017){Siudek}, {Ma{\l}ek}, {Scodeggio}, {Garilli},
  {Pollo}, {Haines}, {Fritz}, {Bolzonella}, {de la Torre}, {Granett}, {Guzzo},
  {Abbas}, {Adami}, {Bottini}, {Cappi}, {Cucciati}, {De Lucia}, {Davidzon},
  {Franzetti}, {Iovino}, {Krywult}, {Le Brun}, {Le F{\`e}vre}, {Maccagni},
  {Marchetti}, {Marulli}, {Polletta}, {Tasca}, {Tojeiro}, {Vergani},
  {Zanichelli}, {Arnouts}, {Bel}, {Branchini}, {Ilbert}, {Gargiulo},
  {Moscardini}, {Takeuchi}, \& {Zamorani}}]{siu17}
{Siudek}, M., {Ma{\l}ek}, K., {Scodeggio}, M., {et~al.} 2017, \aap, 597, A107

\bibitem[{{Smercina} {et~al.}(2018){Smercina}, {Smith}, {Dale}, {French},
  {Croxall}, {Zhukovska}, {Togi}, {Bell}, {Crocker}, {Draine}, {Jarrett},
  {Tremonti}, {Yang}, \& {Zabludoff}}]{sme18}
{Smercina}, A., {Smith}, J.~D.~T., {Dale}, D.~A., {et~al.} 2018, \apj, 855, 51

\bibitem[{{Snyder} {et~al.}(2011){Snyder}, {Cox}, {Hayward}, {Hernquist}, \&
  {Jonsson}}]{sny11}
{Snyder}, G.~F., {Cox}, T.~J., {Hayward}, C.~C., {Hernquist}, L., \& {Jonsson},
  P. 2011, \apj, 741, 77

\bibitem[{{Sobral} {et~al.}(2011){Sobral}, {Best}, {Smail}, {Geach},
  {Cirasuolo}, {Garn}, \& {Dalton}}]{sob11}
{Sobral}, D., {Best}, P.~N., {Smail}, I., {et~al.} 2011, \mnras, 411, 675

\bibitem[{{Sommariva} {et~al.}(2012){Sommariva}, {Mannucci}, {Cresci}, {Maiol
  ino}, {Marconi}, {Nagao}, {Baroni}, \& {Grazian}}]{som12}
{Sommariva}, V., {Mannucci}, F., {Cresci}, G., {et~al.} 2012, \aap, 539, A136

\bibitem[{{Springel} {et~al.}(2005){Springel}, {Di Matteo}, \&
  {Hernquist}}]{spr05}
{Springel}, V., {Di Matteo}, T., \& {Hernquist}, L. 2005, \apj, 620, L79

\bibitem[{{Strateva} {et~al.}(2001){Strateva}, {Ivezi{\'c}}, {Knapp},
  {Narayanan}, {Strauss}, {Gunn}, {Lupton}, {Schlegel}, {Bahcall}, {Brinkmann},
  {Brunner}, {Budav{\'a}ri}, {Csabai}, {Castander}, {Doi}, {Fukugita},
  {Gy{\'{o}}ry}, {Hamabe}, {Hennessy}, {Ichikawa}, {Kunszt}, {Lamb}, {McKay},
  {Okamura}, {Racusin}, {Sekiguchi}, {Schneider}, {Shimasaku}, \&
  {York}}]{str01}
{Strateva}, I., {Ivezi{\'c}}, {\v{Z}}., {Knapp}, G.~R., {et~al.} 2001, \aj,
  122, 1861

\bibitem[{{Szomoru} {et~al.}(2011){Szomoru}, {Franx}, {Bouwens}, {van Dokkum},
  {Labb{\'e}}, {Illingworth}, \& {Trenti}}]{szo11}
{Szomoru}, D., {Franx}, M., {Bouwens}, R.~J., {et~al.} 2011, \apj, 735, L22

\bibitem[{{Szomoru} {et~al.}(2013){Szomoru}, {Franx}, {van Dokkum}, {Trenti},
  {Illingworth}, {Labb{\'e}}, \& {Oesch}}]{szo13}
{Szomoru}, D., {Franx}, M., {van Dokkum}, P.~G., {et~al.} 2013, \apj, 763, 73

\bibitem[{{Tacchella} {et~al.}(2016){Tacchella}, {Dekel}, {Carollo},
  {Ceverino}, {DeGraf}, {Lapiner}, {Mandelker}, \& {Primack}}]{tac16}
{Tacchella}, S., {Dekel}, A., {Carollo}, C.~M., {et~al.} 2016, \mnras, 458, 242

\bibitem[{{Tacchella} {et~al.}(2015){Tacchella}, {Carollo}, {Renzini},
  {Schreiber}, {Lang}, {Wuyts}, {Cresci}, {Dekel}, {Genzel}, {Lilly},
  {Mancini}, {Newman}, {Onodera}, {Shapley}, {Tacconi}, {Woo}, \&
  {Zamorani}}]{tac15}
{Tacchella}, S., {Carollo}, C.~M., {Renzini}, A., {et~al.} 2015, Science, 348,
  314

\bibitem[{{Tacchella} {et~al.}(2018){Tacchella}, {Carollo}, {F{\"o}rster
  Schreiber}, {Renzini}, {Dekel}, {Genzel}, {Lang}, {Lilly}, {Mancini},
  {Onodera}, {Tacconi}, {Wuyts}, \& {Zamorani}}]{tacc18}
{Tacchella}, S., {Carollo}, C.~M., {F{\"o}rster Schreiber}, N.~M., {et~al.}
  2018, \apj, 859, 56

\bibitem[{{Tacconi} {et~al.}(2013){Tacconi}, {Neri}, {Genzel}, {Combes},
  {Bolatto}, {Cooper}, {Wuyts}, {Bournaud}, {Burkert}, {Comerford}, {Cox},
  {Davis}, {F{\"o}rster Schreiber}, {Garc{\'{\i}}a-Burillo}, {Gracia-Carpio},
  {Lutz}, {Naab}, {Newman}, {Omont}, {Saintonge}, {Shapiro Griffin}, {Shapley},
  {Sternberg}, \& {Weiner}}]{tac13}
{Tacconi}, L.~J., {Neri}, R., {Genzel}, R., {et~al.} 2013, \apj, 768, 74

\bibitem[{{Tacconi} {et~al.}(2018){Tacconi}, {Genzel}, {Saintonge}, {Combes},
  {Garc{\'\i}a-Burillo}, {Neri}, {Bolatto}, {Contini}, {F{\"o}rster Schreiber},
  {Lilly}, {Lutz}, {Wuyts}, {Accurso}, {Boissier}, {Boone}, {Bouch{\'e}},
  {Bournaud}, {Burkert}, {Carollo}, {Cooper}, {Cox}, {Feruglio}, {Freundlich},
  {Herrera-Camus}, {Juneau}, {Lippa}, {Naab}, {Renzini}, {Salome}, {Sternberg},
  {Tadaki}, {{\"U}bler}, {Walter}, {Weiner}, \& {Weiss}}]{tac18}
{Tacconi}, L.~J., {Genzel}, R., {Saintonge}, A., {et~al.} 2018, \apj, 853, 179

\bibitem[{{Terrazas} {et~al.}(2016){Terrazas}, {Bell}, {Henriques}, {White},
  {Cattaneo}, \& {Woo}}]{ter16}
{Terrazas}, B.~A., {Bell}, E.~F., {Henriques}, B. M.~B., {et~al.} 2016, \apj,
  830, L12

\bibitem[{{Terrazas} {et~al.}(2017){Terrazas}, {Bell}, {Woo}, \&
  {Henriques}}]{ter17}
{Terrazas}, B.~A., {Bell}, E.~F., {Woo}, J., \& {Henriques}, B. M.~B. 2017,
  \apj, 844, 170

\bibitem[{{Thomas} {et~al.}(2011){Thomas}, {Maraston}, \& {Johansson}}]{tho11}
{Thomas}, D., {Maraston}, C., \& {Johansson}, J. 2011, \mnras, 412, 2183

\bibitem[{{Tran} {et~al.}(2003){Tran}, {Franx}, {Illingworth}, {Kelson}, \&
  {van Dokkum}}]{tra03}
{Tran}, K.-V.~H., {Franx}, M., {Illingworth}, G., {Kelson}, D.~D., \& {van
  Dokkum}, P. 2003, \apj, 599, 865

\bibitem[{{Tran} {et~al.}(2004){Tran}, {Franx}, {Illingworth}, {van Dokkum},
  {Kelson}, \& {Magee}}]{tra04}
{Tran}, K.-V.~H., {Franx}, M., {Illingworth}, G.~D., {et~al.} 2004, \apj, 609,
  683

\bibitem[{{Trujillo} {et~al.}(2007){Trujillo}, {Conselice}, {Bundy}, {Cooper},
  {Eisenhardt}, \& {Ellis}}]{tru07}
{Trujillo}, I., {Conselice}, C.~J., {Bundy}, K., {et~al.} 2007, \mnras, 382,
  109

\bibitem[{{Trujillo} {et~al.}(2011){Trujillo}, {Ferreras}, \& {de La
  Rosa}}]{tru11}
{Trujillo}, I., {Ferreras}, I., \& {de La Rosa}, I.~G. 2011, \mnras, 415, 3903

\bibitem[{{van der Wel} {et~al.}(2009){van der Wel}, {Bell}, {van den Bosch},
  {Gallazzi}, \& {Rix}}]{vdw09}
{van der Wel}, A., {Bell}, E.~F., {van den Bosch}, F.~C., {Gallazzi}, A., \&
  {Rix}, H.-W. 2009, \apj, 698, 1232

\bibitem[{{van der Wel} {et~al.}(2012){van der Wel}, {Bell}, {H{\"a}ussler},
  {McGrath}, {Chang}, {Guo}, {McIntosh}, {Rix}, {Barden}, {Cheung}, {Faber},
  {Ferguson}, {Galametz}, {Grogin}, {Hartley}, {Kartaltepe}, {Kocevski},
  {Koekemoer}, {Lotz}, {Mozena}, {Peth}, \& {Peng}}]{vdw12}
{van der Wel}, A., {Bell}, E.~F., {H{\"a}ussler}, B., {et~al.} 2012, \apjs,
  203, 24

\bibitem[{{van der Wel} {et~al.}(2014){van der Wel}, {Franx}, {van Dokkum},
  {Skelton}, {Momcheva}, {Whitaker}, {Brammer}, {Bell}, {Rix}, {Wuyts},
  {Ferguson}, {Holden}, {Barro}, {Koekemoer}, {Chang}, {McGrath},
  {H{\"a}ussler}, {Dekel}, {Behroozi}, {Fumagalli}, {Leja}, {Lundgren},
  {Maseda}, {Nelson}, {Wake}, {Patel}, {Labb{\'e}}, {Faber}, {Grogin}, \&
  {Kocevski}}]{vdw14}
{van der Wel}, A., {Franx}, M., {van Dokkum}, P.~G., {et~al.} 2014, \apj, 788,
  28

\bibitem[{{van der Wel} {et~al.}(2016){van der Wel}, {Noeske}, {Bezanson},
  {Pacifici}, {Gallazzi}, {Franx}, {Mu{\~n}oz-Mateos}, {Bell}, {Brammer},
  {Charlot}, {Chauk{\'e}}, {Labb{\'e}}, {Maseda}, {Muzzin}, {Rix}, {Sobral},
  {van de Sande}, {van Dokkum}, {Wild}, \& {Wolf}}]{vdw16}
{van der Wel}, A., {Noeske}, K., {Bezanson}, R., {et~al.} 2016, \apjs, 223, 29

\bibitem[{{van Dokkum} \& {Franx}(2001)}]{vd01}
{van Dokkum}, P.~G., \& {Franx}, M. 2001, \apj, 553, 90

\bibitem[{{Vergani} {et~al.}(2008){Vergani}, {Scodeggio}, {Pozzetti}, {Iovino},
  {Franzetti}, {Garilli}, {Zamorani}, {Maccagni}, {Lamareille}, {Le F{\`e}vre},
  {Charlot}, {Contini}, {Guzzo}, {Bottini}, {Le Brun}, {Picat}, {Scaramella},
  {Tresse}, {Vettolani}, {Zanichelli}, {Adami}, {Arnouts}, {Bardelli},
  {Bolzonella}, {Cappi}, {Ciliegi}, {Foucaud}, {Gavignaud}, {Ilbert},
  {McCracken}, {Marano}, {Marinoni}, {Mazure}, {Meneux}, {Merighi}, {Paltani},
  {Pell{\`o}}, {Pollo}, {Radovich}, {Zucca}, {Bondi}, {Bongiorno},
  {Brinchmann}, {Cucciati}, {de la Torre}, {Gregorini}, {Perez-Montero},
  {Mellier}, {Merluzzi}, \& {Temporin}}]{ver08}
{Vergani}, D., {Scodeggio}, M., {Pozzetti}, L., {et~al.} 2008, \aap, 487, 89

\bibitem[{{Vergani} {et~al.}(2010){Vergani}, {Zamorani}, {Lilly}, {Lamareille},
  {Halliday}, {Scodeggio}, {Vignali}, {Ciliegi}, {Bolzonella}, {Bondi},
  {Kova{\v{c}}}, {Knobel}, {Zucca}, {Caputi}, {Pozzetti}, {Bardelli},
  {Mignoli}, {Iovino}, {Carollo}, {Contini}, {Kneib}, {Le F{\`e}vre},
  {Mainieri}, {Renzini}, {Bongiorno}, {Coppa}, {Cucciati}, {de la Torre}, {de
  Ravel}, {Franzetti}, {Garilli}, {Kampczyk}, {Le Borgne}, {Le Brun}, {Maier},
  {Pello}, {Peng}, {Perez Montero}, {Ricciardelli}, {Silverman}, {Tanaka},
  {Tasca}, {Tresse}, {Abbas}, {Bottini}, {Cappi}, {Cassata}, {Cimatti},
  {Guzzo}, {Koekemoer}, {Leauthaud}, {Maccagni}, {Marinoni}, {McCracken},
  {Memeo}, {Meneux}, {Oesch}, {Porciani}, {Scaramella}, {Capak}, {Sanders},
  {Scoville}, \& {Taniguchi}}]{ver10}
{Vergani}, D., {Zamorani}, G., {Lilly}, S., {et~al.} 2010, \aap, 509, A42

\bibitem[{{Wang} {et~al.}(2017){Wang}, {Li}, {Xiao}, {Lin}, {Bershady}, {Law},
  {Merrifield}, {Sanchez}, {Riffel}, {Riffel}, \& {Yan}}]{wan17}
{Wang}, E., {Li}, C., {Xiao}, T., {et~al.} 2017, ArXiv e-prints,
  arXiv:1710.07569

\bibitem[{{Wellons} {et~al.}(2015){Wellons}, {Torrey}, {Ma}, {Rodriguez-Gomez},
  {Vogelsberger}, {Kriek}, {van Dokkum}, {Nelson}, {Genel}, {Pillepich},
  {Springel}, {Sijacki}, {Snyder}, {Nelson}, {Sales}, \& {Hernquist}}]{wel15}
{Wellons}, S., {Torrey}, P., {Ma}, C.-P., {et~al.} 2015, \mnras, 449, 361

\bibitem[{{Whitaker} {et~al.}(2012){Whitaker}, {van Dokkum}, {Brammer}, \&
  {Franx}}]{whi12}
{Whitaker}, K.~E., {van Dokkum}, P.~G., {Brammer}, G., \& {Franx}, M. 2012,
  \apjl, 754, L29

\bibitem[{{Wild} {et~al.}(2016){Wild}, {Almaini}, {Dunlop}, {Simpson},
  {Rowlands}, {Bowler}, {Maltby}, \& {McLure}}]{wil16}
{Wild}, V., {Almaini}, O., {Dunlop}, J., {et~al.} 2016, \mnras, 463, 832

\bibitem[{{Wild} {et~al.}(2009){Wild}, {Walcher}, {Johansson}, {Tresse},
  {Charlot}, {Pollo}, {Le F{\`e}vre}, \& {de Ravel}}]{wil09}
{Wild}, V., {Walcher}, C.~J., {Johansson}, P.~H., {et~al.} 2009, \mnras, 395,
  144

\bibitem[{{Williams} {et~al.}(2017){Williams}, {Giavalisco}, {Bezanson},
  {Cappelluti}, {Cassata}, {Liu}, {Lee}, {Tundo}, \& {Vanzella}}]{wil17}
{Williams}, C.~C., {Giavalisco}, M., {Bezanson}, R., {et~al.} 2017, \apj, 838,
  94

\bibitem[{{Williams} {et~al.}(2010){Williams}, {Quadri}, {Franx}, {van Dokkum},
  {Toft}, {Kriek}, \& {Labb{\'e}}}]{wil10}
{Williams}, R.~J., {Quadri}, R.~F., {Franx}, M., {et~al.} 2010, \apj, 713, 738

\bibitem[{{Willmer} {et~al.}(2006){Willmer}, {Faber}, {Koo}, {Weiner},
  {Newman}, {Coil}, {Connolly}, {Conroy}, {Cooper}, {Davis}, {Finkbeiner},
  {Gerke}, {Guhathakurta}, {Harker}, {Kaiser}, {Kassin}, {Konidaris}, {Lin},
  {Luppino}, {Madgwick}, {Noeske}, {Phillips}, \& {Yan}}]{wil06}
{Willmer}, C.~N.~A., {Faber}, S.~M., {Koo}, D.~C., {et~al.} 2006, \apj, 647,
  853

\bibitem[{{Worthey} \& {Ottaviani}(1997)}]{wor97}
{Worthey}, G., \& {Ottaviani}, D.~L. 1997, \apjs, 111, 377

\bibitem[{{Wu}(2018)}]{wu18a}
{Wu}, P.-F. 2018, \mnras, 473, 5468

\bibitem[{{Wu} {et~al.}(2014){Wu}, {Gal}, {Lemaux}, {Kocevski}, {Lubin},
  {Rumbaugh}, \& {Squires}}]{wu14}
{Wu}, P.-F., {Gal}, R.~R., {Lemaux}, B.~C., {et~al.} 2014, \apj, 792, 16

\bibitem[{{Wu} {et~al.}(2015){Wu}, {Kudritzki}, {Tully}, \& {Neill}}]{wu15}
{Wu}, P.-F., {Kudritzki}, R.-P., {Tully}, R.~B., \& {Neill}, J.~D. 2015, \apj,
  810, 151

\bibitem[{{Wu} {et~al.}(2018){Wu}, {van der Wel}, {Gallazzi}, {Bezanson},
  {Pacifici}, {Straatman}, {Franx}, {Bari{\v s}i{\'c}}, {Bell}, {Brammer},
  {Calhau}, {Chauke}, {van Houdt}, {Maseda}, {Muzzin}, {Rix}, {Sobral},
  {Spilker}, {van de Sande}, {van Dokkum}, \& {Wild}}]{wu18b}
{Wu}, P.-F., {van der Wel}, A., {Gallazzi}, A., {et~al.} 2018, \apj, 855, 85

\bibitem[{{Xu} {et~al.}(2012){Xu}, {Zhao}, {Scoville}, {Capak}, {Drory}, \&
  {Gao}}]{xu12}
{Xu}, C.~K., {Zhao}, Y., {Scoville}, N., {et~al.} 2012, \apj, 747, 85

\bibitem[{{Yan} {et~al.}(2006){Yan}, {Newman}, {Faber}, {Konidaris}, {Koo}, \&
  {Davis}}]{yan06}
{Yan}, R., {Newman}, J.~A., {Faber}, S.~M., {et~al.} 2006, \apj, 648, 281

\bibitem[{{Yan} {et~al.}(2009){Yan}, {Newman}, {Faber}, {Coil}, {Cooper},
  {Davis}, {Weiner}, {Gerke}, \& {Koo}}]{yan09}
---. 2009, \mnras, 398, 735

\bibitem[{{Yang} {et~al.}(2008){Yang}, {Zabludoff}, {Zaritsky}, \&
  {Mihos}}]{yan08}
{Yang}, Y., {Zabludoff}, A.~I., {Zaritsky}, D., \& {Mihos}, J.~C. 2008, \apj,
  688, 945

\bibitem[{{Yano} {et~al.}(2016){Yano}, {Kriek}, {van der Wel}, \&
  {Whitaker}}]{yan16}
{Yano}, M., {Kriek}, M., {van der Wel}, A., \& {Whitaker}, K.~E. 2016, \apjl,
  817, L21

\bibitem[{{Yesuf} {et~al.}(2014){Yesuf}, {Faber}, {Trump}, {Koo}, {Fang},
  {Liu}, {Wild}, \& {Hayward}}]{yes14}
{Yesuf}, H.~M., {Faber}, S.~M., {Trump}, J.~R., {et~al.} 2014, \apj, 792, 84

\bibitem[{{Zahid} \& {Geller}(2017)}]{zah17}
{Zahid}, H.~J., \& {Geller}, M.~J. 2017, \apj, 841, 32

\bibitem[{{Zanella} {et~al.}(2016){Zanella}, {Scarlata}, {Corsini}, {Bedregal},
  {Dalla Bont{\`a}}, {Atek}, {Bunker}, {.~Colbert}, {Dai}, {Henry}, {Malkan},
  {Martin}, {Rafelski}, {Rutkowski}, {Siana}, \& {Teplitz}}]{zan16}
{Zanella}, A., {Scarlata}, C., {Corsini}, E.~M., {et~al.} 2016, \apj, 824, 68

\bibitem[{{Zolotov} {et~al.}(2015){Zolotov}, {Dekel}, {Mandelker}, {Tweed},
  {Inoue}, {DeGraf}, {Ceverino}, {Primack}, {Barro}, \& {Faber}}]{zol15}
{Zolotov}, A., {Dekel}, A., {Mandelker}, N., {et~al.} 2015, \mnras, 450, 2327

\end{thebibliography}

\appendix
\section{Using different definitions of quiescence}
\label{app}

\begin{figure*}
	\includegraphics[width=\textwidth]{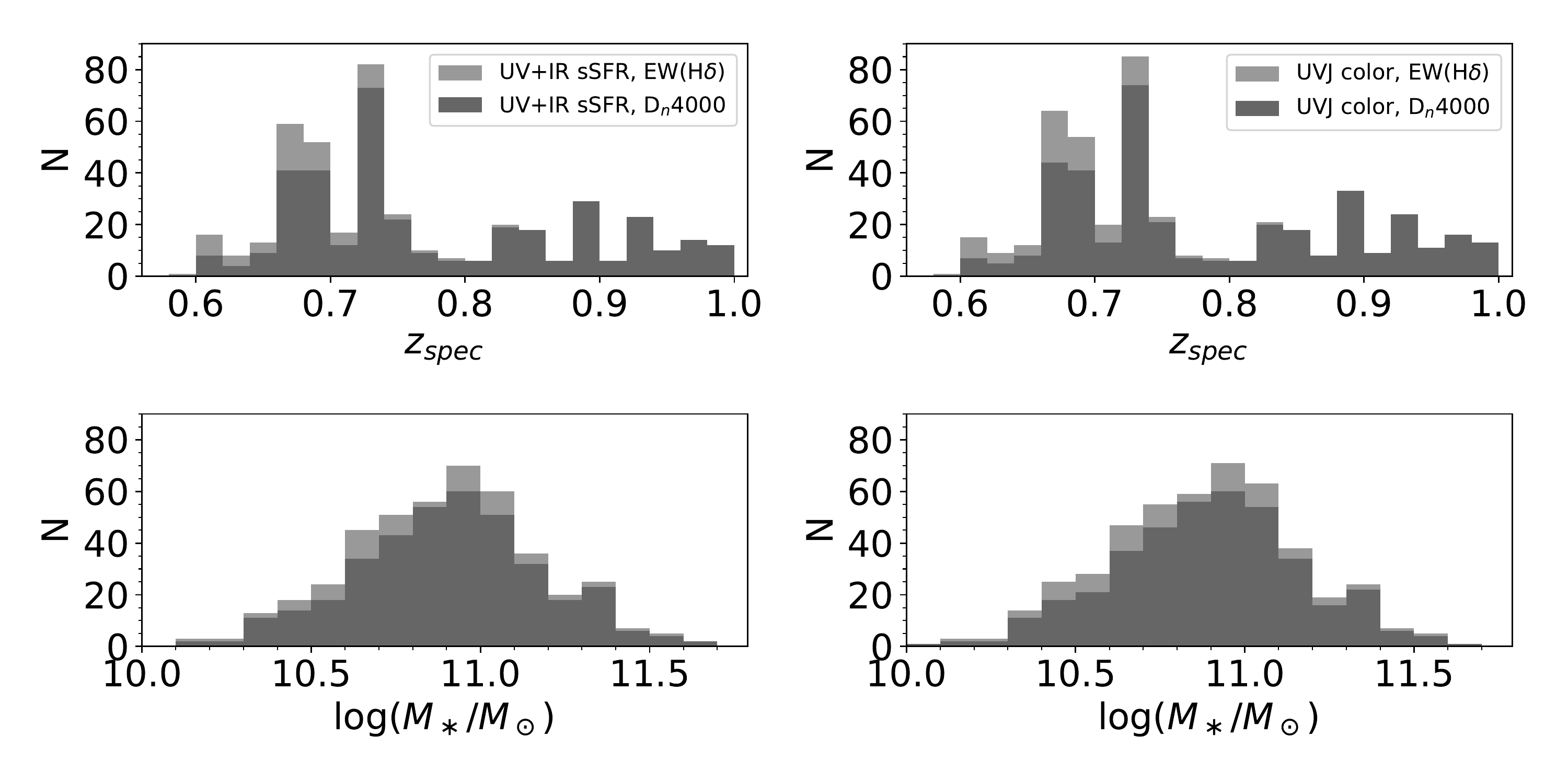}
	\caption{The distributions of redshifts and stellar masses of quiescent galaxies selected based on the sSFR and UVJ color, the same as Fig.~\ref{fig:mz}. The light and dark gray histograms are the distributions of galaxies with only EW(H$\delta$) and both EW(H$\delta$) and D$_n$4000 measurements, respectively. These two selections contains more galaxies at $z \gtrsim 0.8$ comparing to the EW(H$\beta$) selection.}
	\label{fig:mzapp}
\end{figure*}

\begin{figure*}
	\centering
	\includegraphics[width=\textwidth]{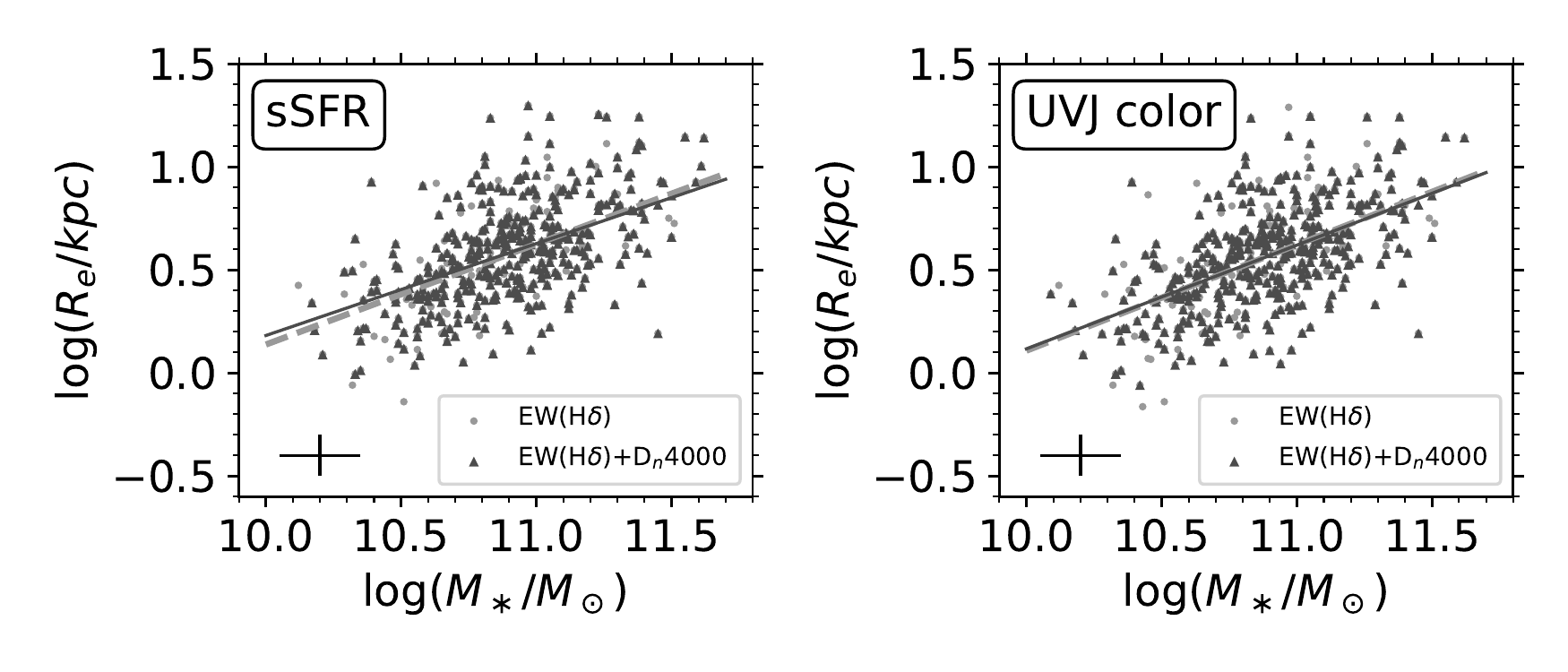}
	\caption{The size and stellar masses of quiescent galaxies selected based on the sSFR and UVJ color, the same as Fig.~\ref{fig:MR}. Dark gray triangles are galaxies whose spectra cover both D$_n$4000 and EW(H$\delta$). Light gray circles are galaxies for which only EW(H$\delta$) is measurable. The light gray dashed line and the dark gray solid line are the best-fit mass-size relation for all galaxies and galaxies with D$_n$4000 measurements, respectively.}
	\label{fig:MRapp}
\end{figure*}

We repeat the analysis in Section~\ref{sec:res} using quiescent galaxies selected by the sSFR and UVJ colors. 
Fig.~\ref{fig:mzapp} show the distributions of redshifts and stellar masses of each sample. Comparing to the EW(H$\beta$) selection used in the main text, these two samples contains galaxies with $z\gtrsim0.8$. The best-fit parameters of the mass-size relation (Fig.~\ref{fig:MRapp}) is listed in Table~\ref{tab:RMapp}. 

\begin{figure*}
	\centering
	\includegraphics[width=\textwidth]{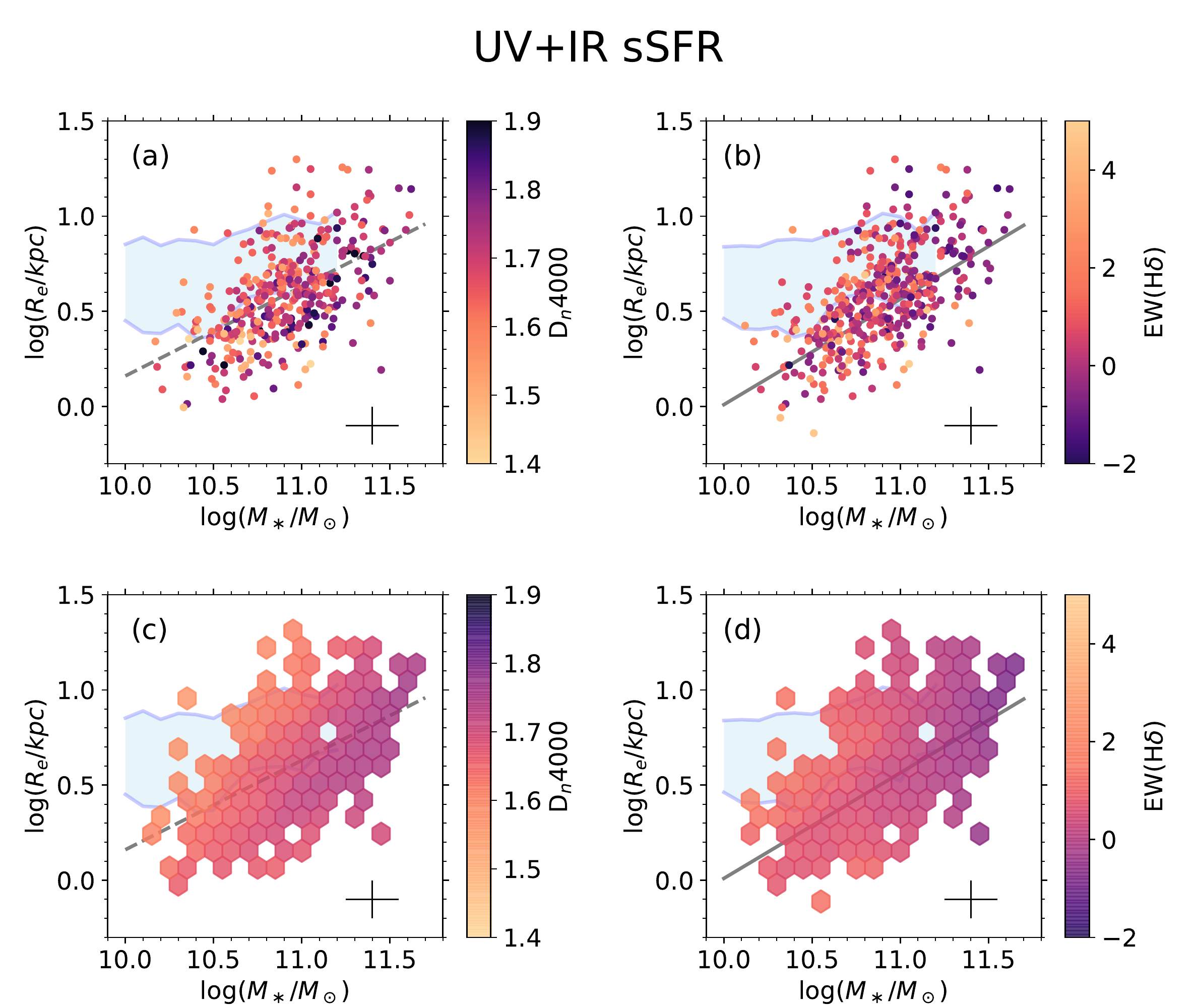}
	\caption{The mass-size relation of quiescent galaxies selected by the sSFR, color-coded by D$_n$4000 and EW(H$\delta$) of individual galaxies and averaging over nearby data points on the mass-size plane using the LOESS method. The light-blue shaded areas are the 16th and 84th percentiles of the sizes of star-forming galaxies at fixed mass. The dashed and solid lines are the best-fit mass-size relations of quiescent galaxies. The cross in the bottom right corner represents for the uncertainties in stellar mass (0.15~dex) and sizes (0.10~dex). }
	\label{fig:MRloessUVIR}
\end{figure*}

\begin{figure*}
	\centering
	\includegraphics[width=\textwidth]{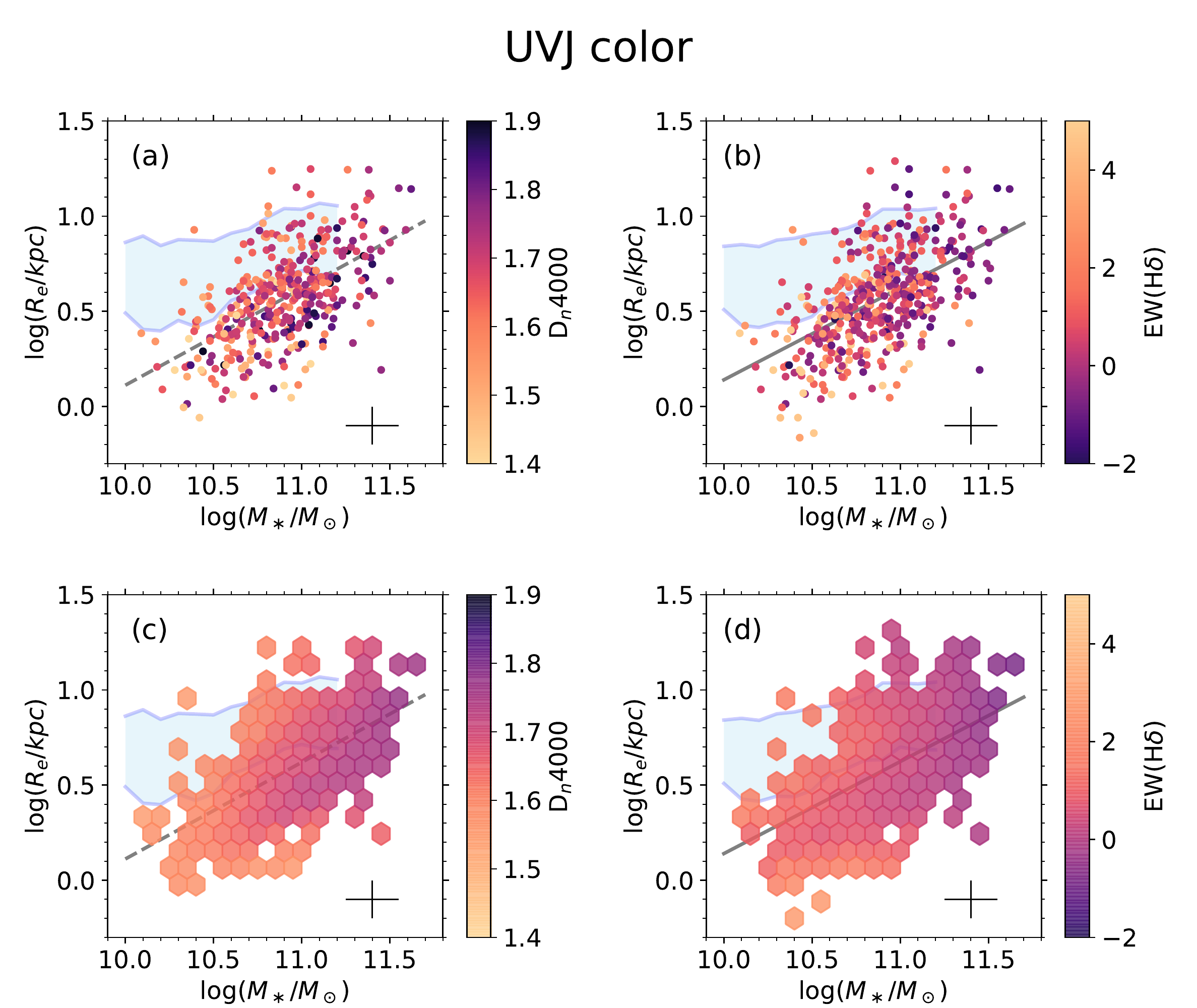}
	\caption{The mass-size relation of quiescent galaxies selected by the UVJ colors, color-coded by D$_n$4000 and EW(H$\delta$) of individual galaxies and averaging over nearby data points on the mass-size plane using the LOESS method. The dependence on the size is not clear. The light-blue shaded areas are the 16th and 84th percentiles of the sizes of star-forming galaxies at fixed mass. The dashed and solid lines are the best-fit mass-size relations of quiescent galaxies.}
	\label{fig:MRloessUVJ}
\end{figure*}

\begin{figure*}
	\centering
	\includegraphics[width=0.9\textwidth]{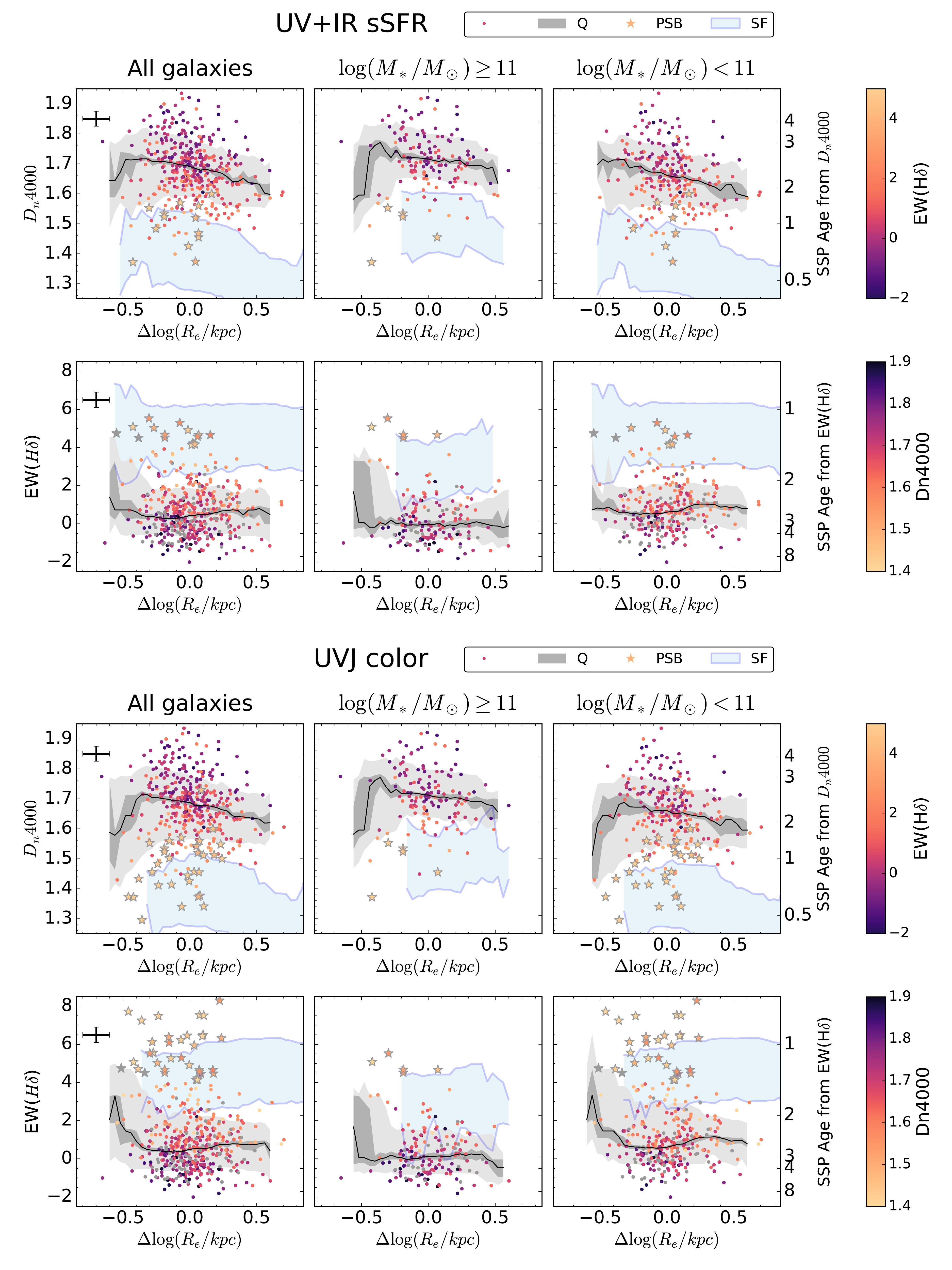}
	\caption{The correlation between the size and the absorption line indices D$_n$4000 and EW(H$\delta$) of quiescent galaxies selected by the sSFR and UVJ colors. The symbols are the same as Fig.~\ref{fig:AR}. At fixed $\log \Delta (R_e)$, there are large variations in D$_n$4000 and EW(H$\delta$) among individual galaxies. Except the most compact galaxies, the median D$_n$4000 is smaller for large galaxies, but the EW(H$\delta$) does not show a clear size dependence. }
	\label{fig:ARapp}
\end{figure*}

\begin{figure*}
	\centering
	\includegraphics[width=0.9\textwidth]{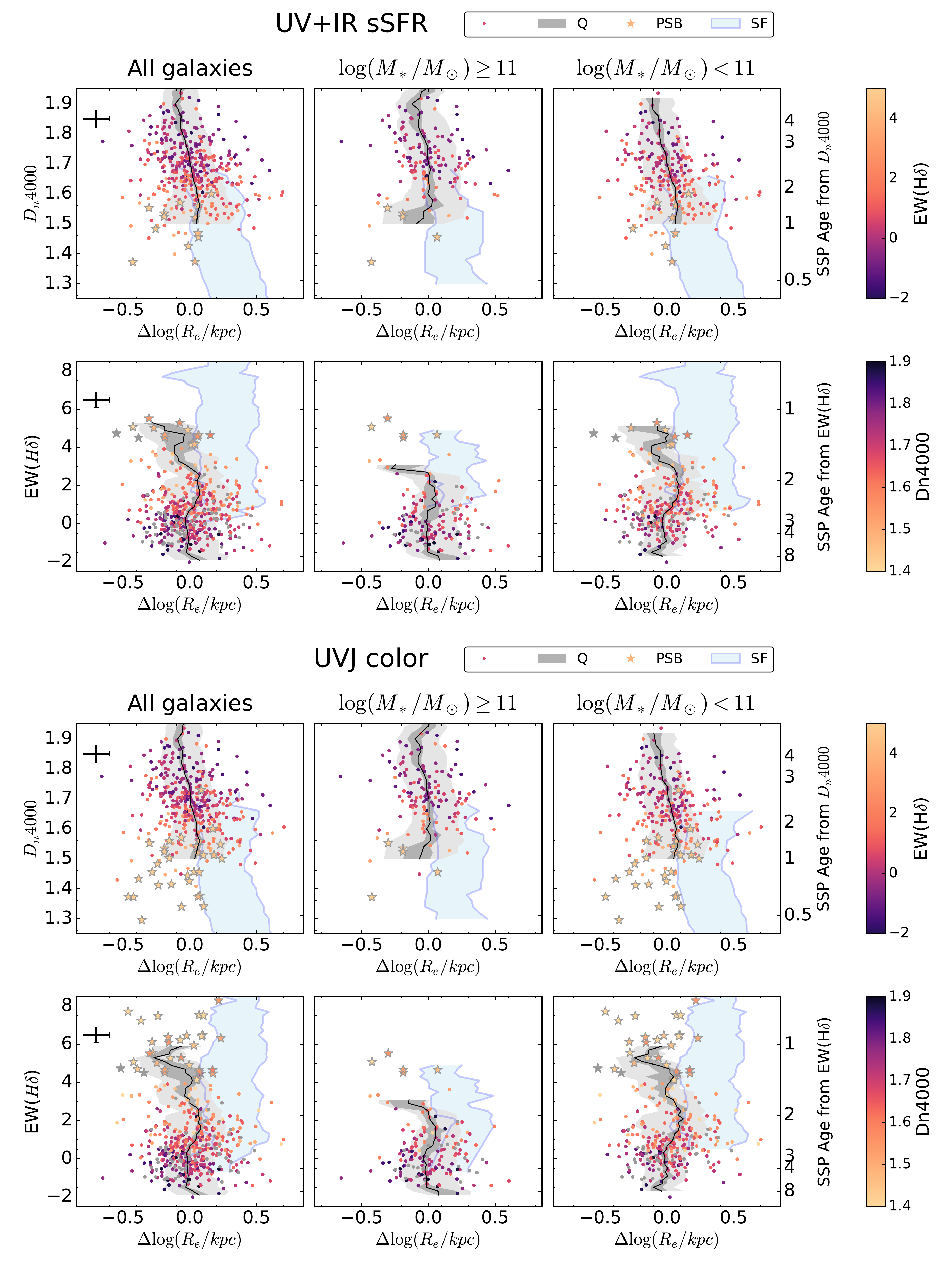}
	\caption{The sizes of quiescent galaxies as a function of D$_n$4000 and EW(H$\delta$), selected by the sSFR and UVJ colors. The symbols are the same Fig.~\ref{fig:RA}. PSB galaxies are on average smaller than other quiescent galaxies and significantly smaller than star-forming galaxies with similar indices. The median $\log \Delta (R_e)$ of the sSFR and UVJ selected PSB galaxies are $-0.07\pm0.08$ and $-0.06\pm0.07$, respectively.}
	\label{fig:RAapp}
\end{figure*}

\begin{figure*}
	\includegraphics[width=\textwidth]{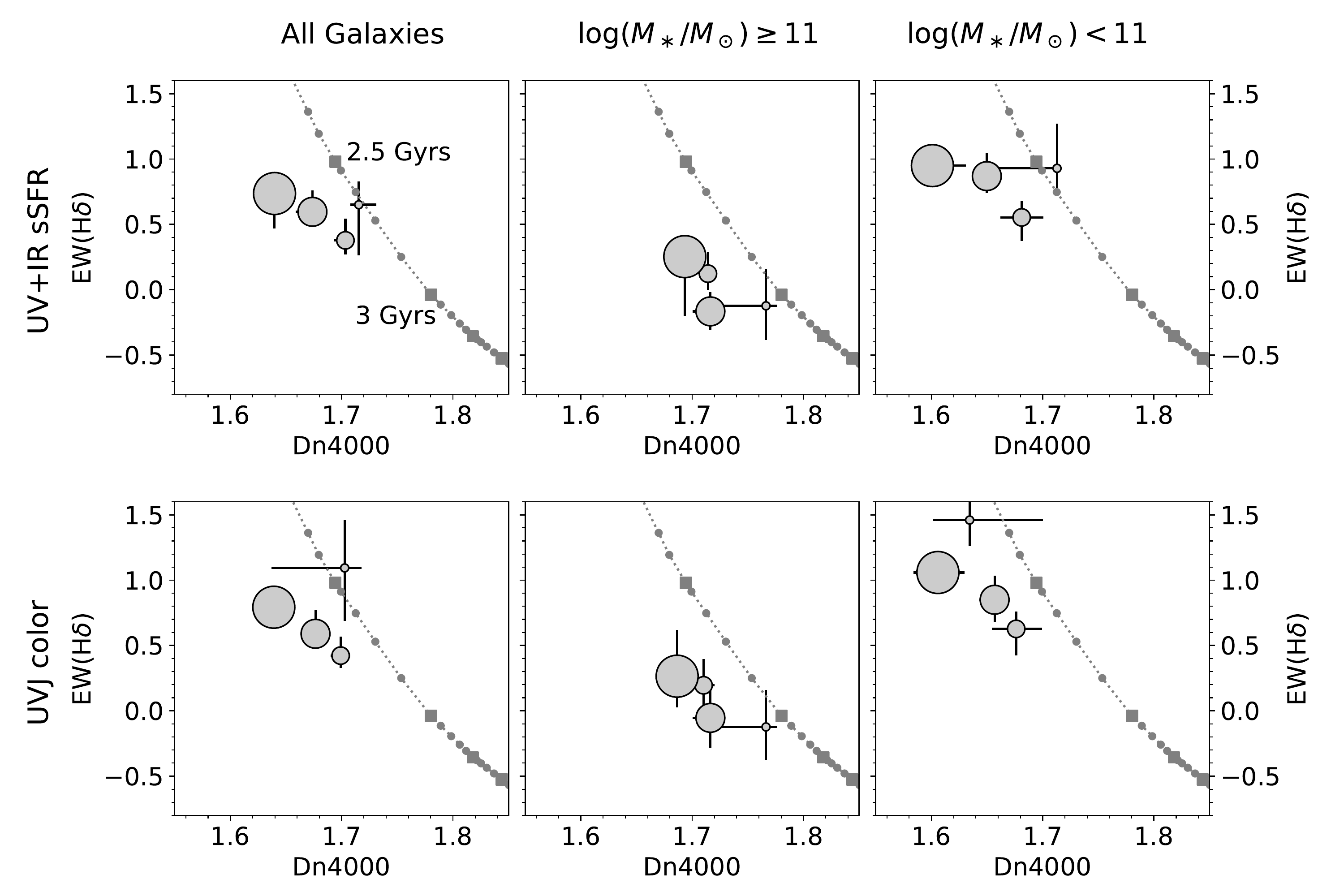}
	\caption{The median D$_n$4000 and EW(H$\delta$) of quiescent galaxies selected base on sSFR and UVJ colors in 4 size bins: $\Delta \log(R_e) > 0.2$, $0.2 \geq \Delta \log(R_e) > 0$, $0 \geq \Delta \log(R_e) > -0.2$, and $-0.2 \geq \Delta \log(R_e)$, the same as in Fig.~\ref{fig:ARDnHd}. Larger circles represent larger galaxies. The uncertainties of medians are calculated from 1000 bootstrap samples. The dotted lines are galaxy evolutionary model tracks for an SSP with solar metallicity. Squares mark the age of 2.5, 3.0, 3.5, and 4.0 Gyrs. Small gray dots label each of 0.1~Gyr time stamp. The 3 larger bins show a weak correlation between the ages and the sizes; larger galaxies are on average younger. There is no such a correlation for massive galaxies.  }
	\label{fig:ARDnHdapp}
\end{figure*}

\begin{figure*}
	\includegraphics[width=\textwidth]{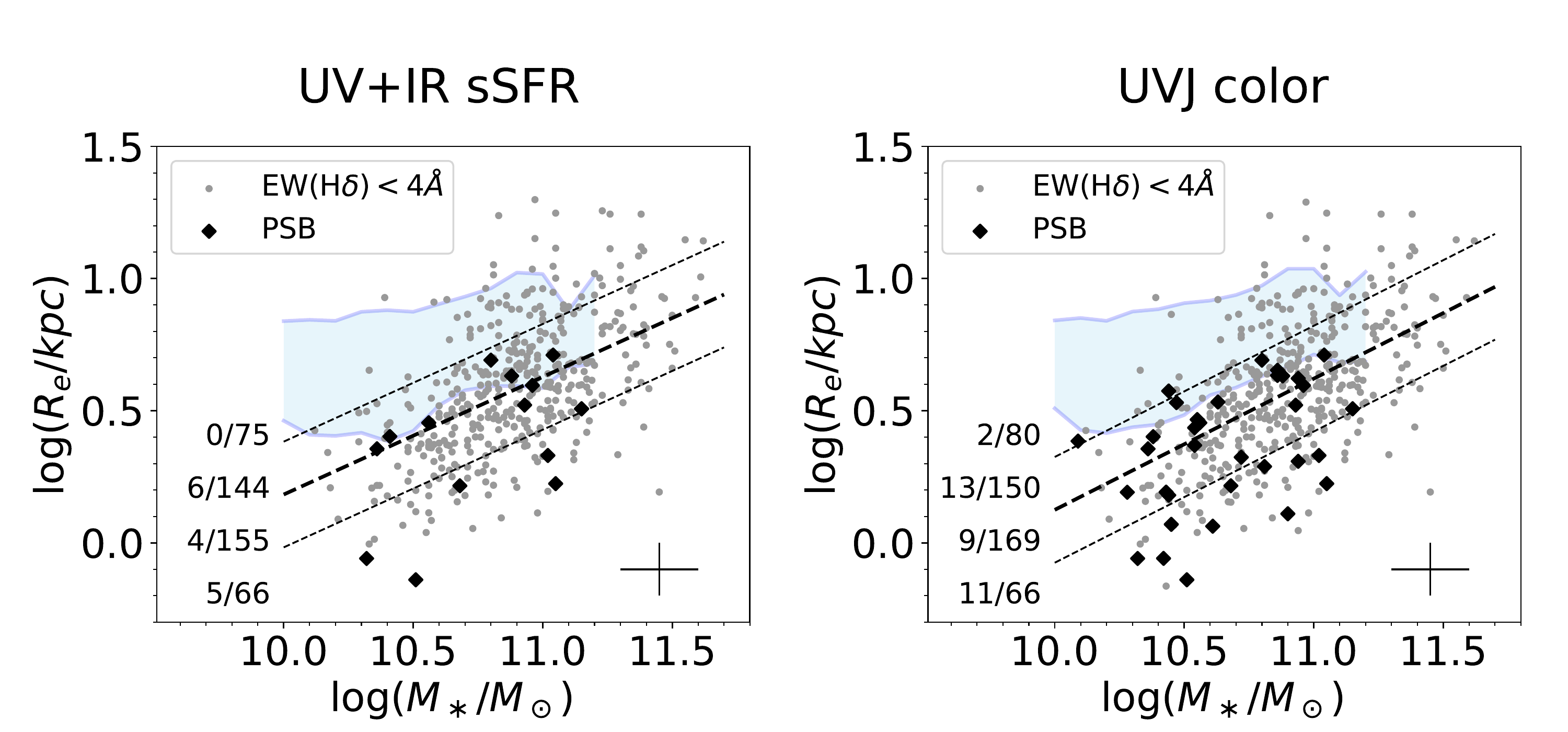}
	\caption{The sizes of PSB galaxies. The `quiescence' is defined by the sSFR or UVJ colors. The symbols are the same as Fig.~\ref{fig:PSB}. The numbers of all quiescent galaxies and PSB galaxies in 4 different ranges of sizes are listed at the left. Most of PSB galaxies are much smaller than star-forming galaxies.}
	\label{fig:PSBapp}
\end{figure*}

We find qualitative the same conclusions using different definitions of quiescence. More massive galaxies have larger D$_n$4000 and smaller EW(H$\delta$) (Fig.~\ref{fig:MRloessUVIR}, \ref{fig:MRloessUVJ}). Despite large individual variations, larger quiescent galaxies have on average smaller D$_n$4000 and larger EW(H$\delta$), indicating younger stellar ages (Fig.~\ref{fig:ARapp}, \ref{fig:ARDnHdapp}). On the other hand, PSB galaxies are not large quiescent galaxies and much smaller than star-forming galaxies (Fig.~\ref{fig:RAapp}, \ref{fig:PSBapp}).

\begin{table}
	\caption{Best-fit Mass-size relation}
	\label{tab:RMapp}
	\begin{center}
		\begin{threeparttable}
			\begin{tabular}{lcc}
				\hline
				\hline
				Selection & a & b \\
				\hline
				UV+IR sSFR & $0.45^{+0.05}_{-0.04}$ & $0.63^{+0.01}_{-0.01}$ \\
				UV+IR sSFR, D$_n$4000 & $0.46^{+0.06}_{-0.06}$ & $0.63^{+0.00}_{-0.01}$ \\
				UVJ color & $0.48^{+0.05}_{-0.06}$ & $0.62^{+0.00}_{-0.00}$ \\
				UVJ color , D$_n$4000 &	$0.50^{+0.04}_{-0.05}$ & $0.62^{+0.00}_{-0.01}$ \\
				\hline
			\end{tabular}
			\begin{tablenotes}
				\small
				\item $\log(R_e) = a \times [\log(M_\ast/M_\odot)-11] + b$
			\end{tablenotes}
		\end{threeparttable}
	\end{center}
\end{table}

\end{document}